\documentclass{optica-article}

\journal{opticajournal} % for journals or Optica Open

\articletype{Research Article}

\usepackage{soul}
\usepackage{lineno}
\linenumbers % Turn off line numbering for Optica Open preprint submissions.
\usepackage{graphicx}
\usepackage{subcaption}
\usepackage{booktabs}
\usepackage{pgfplots}
\usepackage{dblfloatfix}
\pgfplotsset{compat=1.18}
\usepackage{helvet}
\usepackage{tabularx}
\usepackage{makecell}
\usepackage{bm}
\usepackage[table]{xcolor}

\begin{document}
\nolinenumbers

\title{Beyond the Thin-Layer Limit: Differentiable Volumetric Training for Visible-Range Diffractive Neural Networks}

\author{Dineth Jayakody\authormark{1} and Dushan N. Wadduwage\authormark{1,2,3,*}}

\address{\authormark{1}Department of Computer Science, Old Dominion University, Norfolk, VA 23529, USA\\
\authormark{2}School of Data Science, Old Dominion University, Norfolk, VA 23529, USA\\
\authormark{3}Department of Physics, Old Dominion University, Norfolk, VA 23529, USA\\}

\email{\authormark{*}dwadduwa@odu.edu} 

\begin{abstract*}
Diffractive deep neural networks ($D^2$NNs) promise miniaturized power-efficient, light-speed optical front-ends for machine vision, yet the most mature demonstrations remain in the terahertz regime, built from readily fabricated millimeter-scale neurons. Translating $D^2$NNs to the visible range---where nearly all vision pipelines operate---was long blamed on the difficulty of fabricating nanoscale neurons; but even after recent advances removed that barrier, visible-range $D^2$NNs matching their terahertz counterparts remain out of reach. We identify the true obstacle as the thin-layer approximation underlying nearly all $D^2$NN training, which treats each diffractive layer as an infinitely thin mask. It fails not because of the short wavelength, as is commonly assumed, but because the low-refractive-index materials ($n \approx 1.3$--$1.5$) used at visible wavelengths require relief structures thick enough that intra-layer diffraction and phase accumulation become significant. To overcome this, we introduce a differentiable beam-propagation ($\partial$BPM) layer that models each element as a finite-thickness volume and propagates light through it during training, keeping the fabrication-compatible height map end-to-end trainable without full-wave simulation in the loop. Across MNIST, Fashion-MNIST, and CIFAR-100 classification and imaging, $\partial$BPM training substantially reduces the design-to-device mismatch, and full-wave FDTD validation raises classification accuracy from $50\%$ to $90\%$ without re-optimization. The $\partial$BPM layer thus offers a scalable, physics-aware bridge between efficient optical neural-network optimization and fabrication-consistent diffractive design.
\end{abstract*}

\section{Introduction}

\begin{figure}
    \centering
    \includegraphics[width=1\linewidth]{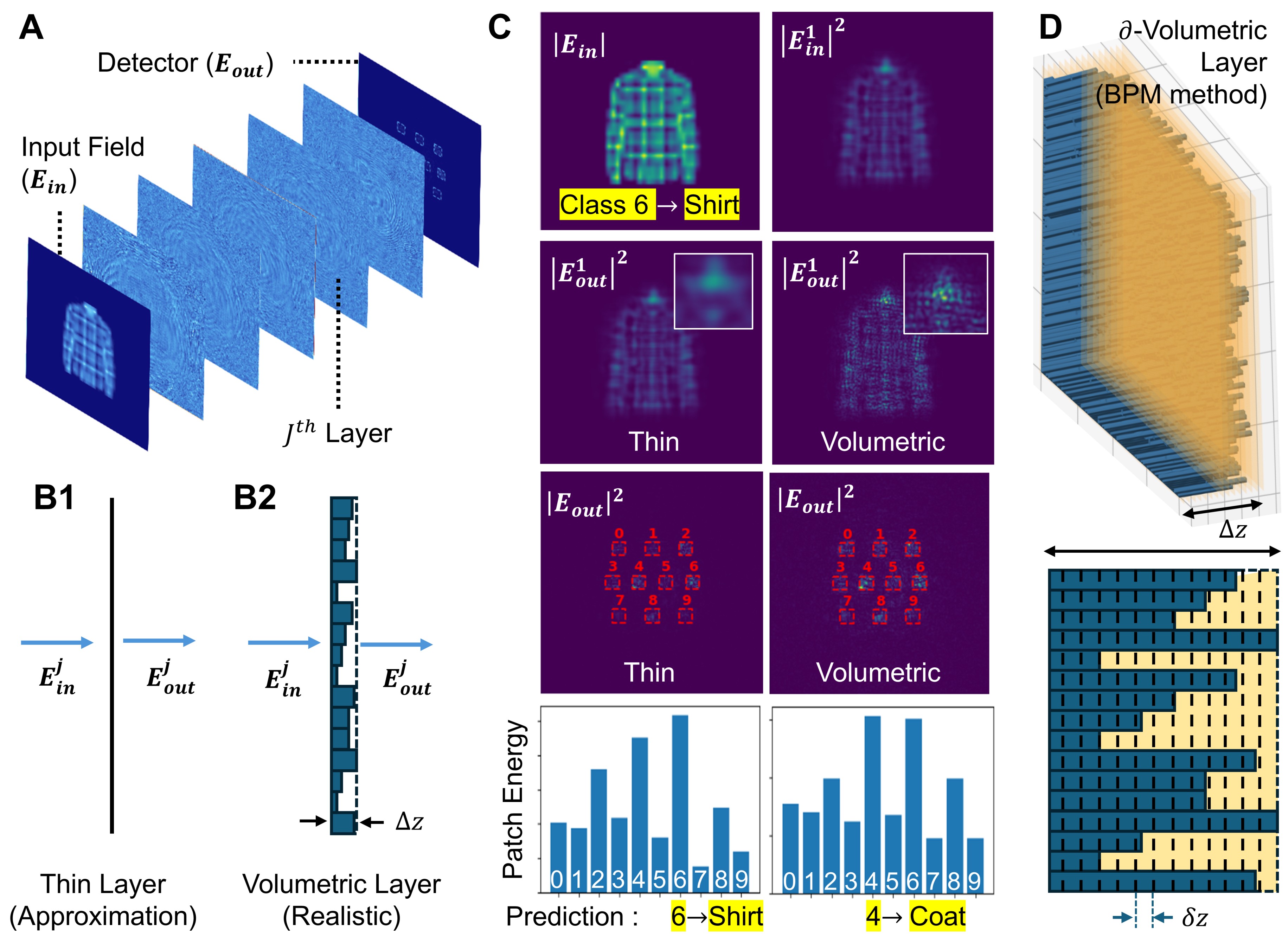}
    \caption{\textbf{The model mismatch issue due to the thin-layer approximation and the proposed differentiable volumetric layer based on BPM.} (A) A schematic of a D2NN with an input field, five thin layers, and the field at the detector plane. (B1) A schematic of the thin-layer approximation used during training. (B2) A schematic of a realistic finite-thickness volumetric layer. (C) Representative fields at the input, immediately after layer 1, and at the output. When the thin approximation is inaccurate, the finite-thickness layer can alter the propagated field enough to change the classification result. (D) A schematic of the proposed differentiable volumetric layer based on BPM.}
    \label{fig:combined_images_fig1}
\end{figure}

Diffractive deep neural networks ($D^2$NNs) optically encode learned transformations into cascaded diffractive layers to perform inference through coherent wave propagation. Instead of executing matrix operations electronically, a $D^2$NN modulates the input optical field as it propagates through a sequence of passive optical elements, forming an output intensity distribution that represents the network decision. Since the original demonstration of all-optical machine learning, $D^2$NNs have been explored as a platform for optical classification, imaging, sensing, machine vision, and communications~\cite{Lin2018Science,Fu2024ONNReview,Chen2024APRReview,Hu2024FreeSpacePerspective}. Interestingly, many early and influential free-space $D^2$NN demonstrations are in the terahertz regime. These models utilize millimeter-scale optical neurons that can be 3D-printed on dielectric materials with reported refractive indices typically in the range of $n\approx1.65$--$1.72$~\cite{Lin2018Science,Luo2019BroadbandLSA,Veli2021THzPulseShaping,Rahman2023OcclusionCommunication,Mengu2023Multispectral,Isil2024Denoising,Li2024ComplexField}. Thus, terahertz $D^2$NNs provide a practical starting point for passive free-space optical inference, enabling diffractive networks to be physically realized with comparatively large feature sizes and experimentally manageable optical layouts.

Despite this progress, $D^2$NNs at visible wavelengths remain comparatively less developed than their terahertz counterparts. This gap is important because visible-light operation would make diffractive optical computing directly compatible with optical systems that already acquire, process, or interpret information in the visible band. For example, Hu et al.~\cite{Hu2024FreeSpacePerspective} identified machine vision, computational imaging, sensing, and optical telecommunications as major application directions for free-space diffractive optical computing. Huang et al.~\cite{Huang2024PreSensor} further showed that optical neural networks can perform pre-sensor computation before image capture, supporting the idea that part of the visual processing pipeline can be moved into the optical domain. Diffractive processors have also been used for: single-pixel machine vision by optically encoding spatial information into the output spectrum~\cite{Li2021SinglePixel}; all-optical image denoising before digital reconstruction~\cite{Isil2024Denoising}; and complex-field imaging that retrieves amplitude and quantitative phase information using an intensity-based sensor without digital phase retrieval~\cite{Li2024ComplexField}. $D^2$NNs have also been used for non-imaging tasks. For instance, $D^2$NNs could counteract wavefront distortions~\cite{Shen2024PhaseConjugation} using all-optical phase conjugation. Together, these works show that visible-compatible diffractive optical learning could be valuable for front-end optical computation in imaging, sensing, and machine-vision pipelines, where information is often already carried by optical fields before electronic detection. Despite this potential, visible-range $D^2$NNs remain less experimentally mature than their terahertz counterparts because visible-light operation requires much smaller, often micro- to nanoscale, diffractive features, making fabrication, alignment, and physically accurate modeling substantially more challenging.

In principle, a $D^2$NN designed for terahertz wavelengths can be translated to visible wavelengths by simply downscaling all its dimensions by the ratio of the wavelengths. However, the resulting design requires nanometer-scale optical neurons, creating a substantial fabrication difficulty. Nevertheless, in their seminal work, Chen et al.~\cite{Chen2021Visible} demonstrated the first $D^2$NNs in the visible range through an elegant compromise. The maximum half-cone diffraction angle of a neuron is given by $u_{\max}=\sin^{-1}(\lambda/2d_f)$, where $\lambda$ is the operating wavelength and $d_f$ is the neuron size. In the ideal geometric limit, choosing $d_f$ close to $\lambda/2$ maximizes the propagating diffraction angle of each neuron.
 With low diffraction angles from larger neurons, light from one neuron spreads over a smaller region of the next diffractive layer, reducing the connections between layers. To regain the connectivity, Chen et al.~\cite{Chen2021Visible} proposed to increase the distance between layers allowing the same effective connectivity but with much larger micrometer-scale neurons. For a fully connected configuration, the lateral diffraction radius after an inter-layer spacing $D$, $R=D\tan(u_{\max})$, should be comparable to or larger than the side length of the square layer, $w=\sqrt{N}d_f$ (see Supplementary Sections~3 and~4). Therefore, visible-wavelength $D^2$NNs face a tradeoff among wavelength, neuron size, number of neurons, and layer spacing. Chen et al.~\cite{Chen2021Visible} followed the larger-spacing route by using micrometer-scale neurons and centimeter-scale inter-layer distances, enabling visible-light operation but at the cost of a larger optical footprint. Recent nanoprecise fabrication advances have begun to address this miniaturization bottleneck. Recently, Yang et al.~\cite{yang2026isotropic} introduced implosion carving (ImpCarv), a hydrogel-based method in which vacancies are photopatterned inside a swollen scaffold and then isotropically shrunk to produce three-dimensional refractive-index structures with nanoscale precision. This approach enabled visible-wavelength $D^2$NN with nanoscale neuron sizes, using neurons of approximately $500~\mathrm{nm}\times500~\mathrm{nm}$ and operation at $\lambda=532~\mathrm{nm}$. Interestingly, although ImpCarv is capable of fabricating nanoprecise features, Yang et al.~\cite{yang2026isotropic} did not ultimately adopt half-$\lambda$-sized neurons for their visible-wavelength $D^2$NN device. Although such neurons maximize the available propagating diffraction-angle range, the authors found that, after fabrication, these deeply subwavelength features could deviate strongly from the intended design. They therefore selected a larger neuron size, balancing optical connectivity with more reliable physical implementation.

% Although half-$\lambda$ neurons maximize the available propagating diffraction-angle range, they found that such small neurons can deviate strongly from the intended design in realistic material settings. Therefore, they selected neurons with a size close to $\lambda$, balancing angular spread with more reliable physical implementation.

The challenge originates from a dominant assumption in the $D^2$NN training pipelines: each diffractive layer is treated as an infinitely thin complex transmission mask, as illustrated in Fig.~\ref{fig:combined_images_fig1}(B1). In the standard formulation, the optical field immediately after layer $\ell$ is written as 
\begin{equation}
u_\ell^{+}(x,y)=u_\ell^{-}(x,y)\,t_\ell(x,y),
\end{equation}
where $u_\ell^{-}(x,y)$ and $u_\ell^{+}(x,y)$ denote the complex optical fields immediately before and after the $\ell$-th layer, respectively, and $t_\ell(x,y)$ is a phase-only or complex-valued transmission function. Under this model, propagation is simulated only in the free-space region between adjacent layers~\cite{Lin2018Science,Luo2019BroadbandLSA}. The physical thickness of a fabricated diffractive element does not appear explicitly in the forward model; instead, the learned phase is converted to a height profile after training through a phase-to-thickness relation determined by the wavelength and refractive-index contrast~\cite{Chen2021Visible,Chen2024APRReview}. Yang et al.~\cite{yang2026isotropic} showed that light can propagate and partially confine inside the finite-height neuron, producing effective-index changes and phase errors that are not captured by a local thin-layer model. Of note, this critical limitation had not affected previous THz demonstration with half-$\lambda$ neurons.

% In that model, the layer is treated as an infinitesimally thin transmission layer, where the field is updated as $u^+(x,y)=u^-(x,y)\exp[i\Delta\phi(x,y)]$. Here, $u^-(x,y)$ and $u^+(x,y)$ are the fields before and after the layer, and $\Delta\phi(x,y)$ is the local phase delay. The phase delay is usually related to the physical height by $\Delta\phi(x,y)=2\pi\Delta n h(x,y)/\lambda$, where $\Delta n$ is the refractive-index contrast, $\lambda$ is the wavelength, and $h(x,y)$ is the local height. We attribute this limitation to the absence of intra-layer propagation in the thin-layer forward model. Therefore, this gap can be addressed by introducing a differentiable volumetric forward model that accounts for intra-layer propagation during training, instead of applying the height profile only as a thin-layer approximation.

%Their FDTD analysis showed that light can propagate and partially confine inside the finite-height neuron, producing effective-index changes and phase errors that are not captured by a local thin-layer model.

In this work, we investigate the limitations of the thin-layer approximation and why it affects terahertz vs. visible $D^2$NNs differently. Our results suggest that the mismatch introduced by the thin-layer approximation is not caused by wavelength; it is caused by the low- to moderate-index materials used in visible-range $D^2$NNs. We then address this limitation by introducing an end-to-end differentiable volumetric $D^2$NN layer based on the beam propagation method (BPM). Instead of representing each diffractive layer as a zero-thickness transmission layer, the proposed model converts the learned phase profile into a finite-height structure and propagates the optical field through the axial volume of the layer during training. This allows intra-layer diffraction, cumulative phase accumulation, and thickness-dependent field evolution to influence the optimized design directly. The end-to-end differentiability is important because it allows the finite-thickness geometry to be optimized with standard gradient-based learning, without requiring full-wave simulation at every training step. We evaluate this framework in visible-wavelength and low- to moderate-index regimes relevant to emerging fabrication platforms, and show that volumetric training reduces the model mismatch between thin-layer prediction and finite-thickness propagation. Together, these results position BPM-based volumetric training as a scalable bridge between efficient optical neural-network optimization and more physically consistent finite-thickness diffractive design.

\section{Results and Discussion}
\label{sec:results}

In this section, we first present our proposed differentiable-BPM-layer ($\partial$BPM layer). Next, we present a comprehensive comparison between the $D^2$NNs with the $\partial$BPM layers and the thin layers (as a baseline) on multiple tasks and multiple datasets. Finally, we present a Finite-Difference Time-Domain (FDTD) study to demonstrate the superiority of our  $\partial$BPM layer under the full-wave simulations. 

Across all experiments, we used a common optical backbone so that differences between the $\partial$BPM layers and the thin layers could be attributed to the layer representation rather than unrelated implementation choices. Task-specific variations were limited to the input encoding scheme, output readout strategy, loss function, and optical geometry. Unless otherwise stated, distances and dimensions are expressed in wavelength-normalized units.

\subsection{Differentiable BPM ($\partial$BPM) layer models trainable volumetric intra-layer propagation.}

%We implemented the proposed volumetric formulation as Differentiable BPM-$D^2$NN, a differentiable BPM-based $D^2$NN framework that was used in place of the conventional thin layer during optical network training. Each diffractive layer was parameterized by a trainable phase profile, converted into a fabrication-compatible height map, and then evaluated as a finite-thickness optical volume through a sequence of axial propagation steps. This kept the optimization pipeline compatible with gradient-based learning while allowing the forward pass to depend on the internal axial structure of the layer.

%Across all experiments, we used a common optical backbone so that differences between the thin layer and volumetric models could be attributed to the layer representation rather than unrelated implementation choices. Task-specific variations were limited to the input encoding scheme, output readout strategy, loss function, and optical geometry. Unless otherwise stated, distances and dimensions are expressed in wavelength-normalized units to make the results comparable across operating wavelengths.

%\subsubsection{Differentiable BPM-$D^2$NN forward model}

The central idea of our $\partial$BPM layer is to train a $D^2$NN using a finite-thickness volumetric representation of each diffractive layer, rather than optimizing a zero-thickness thin-layer and converting it to a physical height profile post-training. Our proposed $\partial$BPM layer directly represents a fabrication-compatible height map, evaluated through a differentiable sequence of slice-wise phase modulation and axial propagation steps. Therefore, intra-layer propagation became part of the training forward pass, and gradients could be backpropagated through the full height map during the backward pass. In order to make the height map differentiable, we utilize a differentiable soft occupancy function. 

%In this formulation, each trainable layer was represented as a fabrication-compatible height map, embedded into a finite-thickness optical volume, and evaluated through a differentiable sequence of slice-wise phase modulation and axial propagation steps. Therefore, intra-layer propagation became part of the training forward pass, and gradients could be backpropagated through the full volumetric optical structure.

Let \(U_0(x,y)\) denote the input optical field and let \(\Theta=\{\beta_\ell\}_{\ell=1}^{L}\) denote the trainable parameters of an \(L\)-layer diffractive network. Each layer \(\ell\) was parameterized using trainable latent variables \(\beta_\ell(x,y)\), which were mapped to a wrapped phase profile,
\begin{equation}
\phi_\ell(x,y)=\operatorname{mod}\!\big(2\pi\beta_\ell(x,y),\,2\pi\big),
\qquad
\phi_\ell(x,y)\in[0,2\pi).
\label{eq:wrapped_phase}
\end{equation}
For a material with refractive index \(n\) in air, the index contrast is \(\Delta n=n-1\). The corresponding fabrication-compatible height map was obtained as
\begin{equation}
h_\ell(x,y)=\frac{\lambda}{2\pi\Delta n}\,\phi_\ell(x,y),
\label{eq:height_from_phase_main}
\end{equation}
where \(\lambda\) is the wavelength. This mapping ensured that the learned optical modulation was directly expressed as a finite-thickness physical geometry during training.

Propagation inside the volumetric layer and between diffractive layers was modeled using the angular-spectrum method (ASM). For a propagation distance \(z\), we define the propagation operator \(\mathcal{P}_{z}(\cdot)\) as
\begin{equation}
\mathcal{P}_{z}\!\left[U(x,y;0)\right]
=
\mathcal{F}^{-1}
\left\{
\mathcal{F}\!\left[U(x,y;0)\right]\,
H(f_x,f_y;z)
\right\},
\label{eq:asm_main}
\end{equation}
where \(\mathcal{F}\) and \(\mathcal{F}^{-1}\) denote the 2D Fourier transform and inverse Fourier transform, and \((f_x,f_y)\) are the spatial frequency coordinates corresponding to \((x,y)\). The ASM transfer function is
\begin{equation}
H(f_x,f_y;z)
=
\exp\!\left(
i k_0 z\,\gamma(f_x,f_y)
\right),
\label{eq:asm_transfer}
\end{equation}
with
\begin{equation}
k_0=\frac{2\pi}{\lambda},
\qquad
\gamma(f_x,f_y)=\sqrt{1-(\lambda f_x)^2-(\lambda f_y)^2}.
\label{eq:asm_gamma}
\end{equation}
We evaluated the square-root term using analytic continuation to account for evanescent components:
\begin{equation}
\gamma(f_x,f_y)=
\begin{cases}
\sqrt{1-\lambda^2 f_x^2-\lambda^2 f_y^2}, & 1-\lambda^2 f_x^2-\lambda^2 f_y^2 \ge 0,\\[4pt]
i\sqrt{\lambda^2 f_x^2+\lambda^2 f_y^2-1}, & \text{otherwise}.
\end{cases}
\label{eq:gamma_analytic}
\end{equation}

To evaluate a finite-thickness diffractive layer, we discretized the maximum layer thickness into \(N_{\text{sub}}\) axial sub-slices. The slice thickness was
\begin{equation}
\Delta z = \frac{h_{\max}}{N_{\text{sub}}},
\label{eq:delta_z_sub}
\end{equation}
where
\begin{equation}
h_{\max}=\frac{\lambda}{2\pi\Delta n}\phi_{\max},
\qquad
\phi_{\max}=2\pi.
\label{eq:hmax}
\end{equation}
For slice index \(s=1,\dots,N_{\text{sub}}\), the slice-center depth was
\begin{equation}
z_s=\left(s-\frac{1}{2}\right)\Delta z.
\label{eq:zs}
\end{equation}

The spatial occupancy of material at each axial slice was represented using a differentiable soft occupancy function,
\begin{equation}
m_{\ell,s}(x,y)=
\sigma\!\left(
\frac{h_\ell(x,y)-z_s}{\tau}
\right),
\label{eq:occupancy_main}
\end{equation}
where \(\sigma(\cdot)\) is the sigmoid function and \(\tau\) controls the softness of the slice boundary. This occupancy function approximates whether a spatial location \((x,y)\) contains material at depth \(z_s\), while preserving differentiability with respect to the height map \(h_\ell(x,y)\).

The phase accumulated inside slice \(s\) of layer \(\ell\) was
\begin{equation}
\Delta\phi_{\ell,s}(x,y)
=
k_0 \Delta n \Delta z\,m_{\ell,s}(x,y).
\label{eq:slicephase_main}
\end{equation}
Each volumetric slice then applied a phase delay followed by axial propagation over \(\Delta z\):
\begin{equation}
U
\leftarrow
\mathcal{P}_{\Delta z}
\!\left[
U\exp\!\left(i\Delta\phi_{\ell,s}(x,y)\right)
\right],
\qquad s=1,\dots,N_{\text{sub}},
\label{eq:bpm_layer_main}
\end{equation}
where \(\mathcal{P}_{\Delta z}(\cdot)\) is the ASM propagation operator defined in Eq.~\eqref{eq:asm_main}. Repeating Eq.~\eqref{eq:bpm_layer_main} across all sub-slices produced a differentiable beam-propagation approximation of a finite-thickness diffractive layer.

For a nominal inter-layer spacing \(z_{\text{diff}}\), volumetric propagation through the layer occupied part of the axial distance. Therefore, when a layer was modeled volumetrically, the remaining free-space distance between adjacent diffractive planes was reduced accordingly, as shown in Fig.~\ref{fig:combined_images_fig1}(D). This kept the total axial distance between network planes consistent when comparing the proposed $\partial$BPM layers with the conventional thin-layer baseline.

The complete optical network can be written as
\begin{equation}
U_{\text{out}} = \mathcal{T}_{\Theta}(U_0),
\label{eq:network_operator}
\end{equation}
where \(\mathcal{T}_{\Theta}\) denotes the cascade of volumetric diffractive layers and free-space propagation steps. The output intensity is
\begin{equation}
I_{\text{out}} = U_{\text{out}}U_{\text{out}}^{*}.
\label{eq:output_intensity}
\end{equation}
For a task-specific loss \(\mathcal{L}(I_{\text{out}})\), gradients were propagated through the complete computational graph using automatic differentiation. The differentiable operations included FFT/IFFT-based propagation, complex multiplication, complex exponentiation, sigmoid occupancy, intensity formation, and the task-specific loss.

The differentiability of the volumetric layer follows from the differentiable height-to-occupancy and occupancy-to-phase mappings. From Eq.~\eqref{eq:occupancy_main},
\begin{equation}
\frac{\partial m_{\ell,s}}{\partial h_\ell}
=
\frac{1}{\tau}m_{\ell,s}(1-m_{\ell,s}).
\label{eq:dm_dh}
\end{equation}
From Eq.~\eqref{eq:slicephase_main},
\begin{equation}
\frac{\partial \Delta\phi_{\ell,s}}{\partial m_{\ell,s}}
=
k_0 \Delta n \Delta z.
\label{eq:dphidm}
\end{equation}
For the phase modulation term,
\begin{equation}
\frac{\partial}{\partial \Delta\phi_{\ell,s}}
\left(
Ue^{i\Delta\phi_{\ell,s}}
\right)
=
iUe^{i\Delta\phi_{\ell,s}}.
\label{eq:dexp}
\end{equation}
From the phase-to-height mapping in Eq.~\eqref{eq:height_from_phase_main},
\begin{equation}
\frac{\partial h_\ell}{\partial \phi_\ell}
=
\frac{\lambda}{2\pi\Delta n},
\label{eq:dhdphi}
\end{equation}
and away from the measure-zero phase-wrapping boundaries in Eq.~\eqref{eq:wrapped_phase},
\begin{equation}
\frac{\partial \phi_\ell}{\partial \beta_\ell}=2\pi.
\label{eq:dphidbeta}
\end{equation}
By the chain rule, the gradient with respect to the trainable layer parameters can be expressed as
\begin{equation}
\frac{\partial \mathcal{L}}{\partial \beta_\ell}
=
\sum_{s=1}^{N_{\text{sub}}}
\frac{\partial \mathcal{L}}{\partial U_{\text{out}}}
\frac{\partial U_{\text{out}}}{\partial \Delta\phi_{\ell,s}}
\frac{\partial \Delta\phi_{\ell,s}}{\partial m_{\ell,s}}
\frac{\partial m_{\ell,s}}{\partial h_\ell}
\frac{\partial h_\ell}{\partial \phi_\ell}
\frac{\partial \phi_\ell}{\partial \beta_\ell}.
\label{eq:full_chain}
\end{equation}
Thus, our $\partial$BPM layer enabled end-to-end optimization of fabrication-compatible finite-thickness diffractive layers by allowing gradients to pass through all volumetric slices and free-space propagation steps.

%\subsubsection{Thin-layer baseline}

\paragraph{Thin-layer baseline:} The conventional thin-layer approximation was used as the baseline forward model for comparison. In this formulation, each diffractive layer was treated as an infinitesimally thin transmission layer with no intra-layer propagation. The field update at layer \(\ell\) was written as
\begin{equation}
u_\ell^{+}(x,y)
\leftarrow
u_\ell^{-}(x,y)\,t_\ell(x,y),
\label{eq:thinlayer_main}
\end{equation}
where \(u_\ell^{-}(x,y)\) and \(u_\ell^{+}(x,y)\) denote the fields immediately before and after the layer, respectively, and \(t_\ell(x,y)\) is the transmission function of the \(\ell\)-th layer. In the phase-only setting used here, the transmission function becomes
\begin{equation}
t_\ell(x,y)=\exp\!\left(i\phi_\ell(x,y)\right).
\label{eq:thinlayer_phase_only}
\end{equation}
Unlike the $\partial$BPM layer, this baseline applies the learned phase modulation at a single axial plane and does not model propagation through the physical thickness of the fabricated relief structure. 
%Therefore, it provides a direct comparison between conventional zero-thickness training and the proposed differentiable volumetric training formulation.

\subsection{Differentiable BPM-$D^2$NN learns physically consistent finite-thickness $D^2$NNs for optical classification and imaging.}

%where low- and moderate-index transparent materials make finite-thickness effects especially relevant.

We evaluated the proposed $\partial$BPM-based \(D^2\)NN framework on optical classification and imaging tasks while varying the refractive index of the diffractive material. The refractive index range spanned from 1.1 to 1.9 to include a range of materials used to fabricate \(D^2\)NN. While the main experiments were conducted at \(\lambda=532~\mathrm{nm}\), emphasizing the visible-light regime, the same finite-thickness behavior is expected across wavelength regimes when the geometry is scaled consistently (see supplementary Section~1 for equivalent experiments at \(\lambda=750~\mu\mathrm{m}\), used in the original \(D^2\)NN demonstration~\cite{Lin2018Science}). We present the geometry in wavelength-normalized units, so all key dimensions, including aperture, pixel pitch, layer spacing, and layer thickness, scale with \(\lambda\).

To quantify the effect of modeling fidelity, we compared three settings: (i) thin-layer training and inference, (ii) $\partial$BPM layer training and inference, and (iii) thin-layer training and $\partial$BPM layer inference. The third setting measures the model mismatch between conventional thin-layer training and volumetric evaluation. Specifically, the models were first trained using the standard thin-layer approximation. The learned phase profiles were then converted into physical height maps using the wavelength and refractive-index contrast, as would be done for fabrication. During testing, however, these height profiles were inferred using the BPM forward model, which propagates the field through the finite thickness of each layer. This thin-trained $\rightarrow$ BPM-tested configuration therefore estimates the effect of deploying a conventionally trained thin-layer design under more physically realistic volumetric propagation. We train all models under identical conditions to ensure a fair comparison.

\subsubsection{Results on the classification tasks}

\begin{figure}[htbp]
    \centering
    \includegraphics[width=1\linewidth]{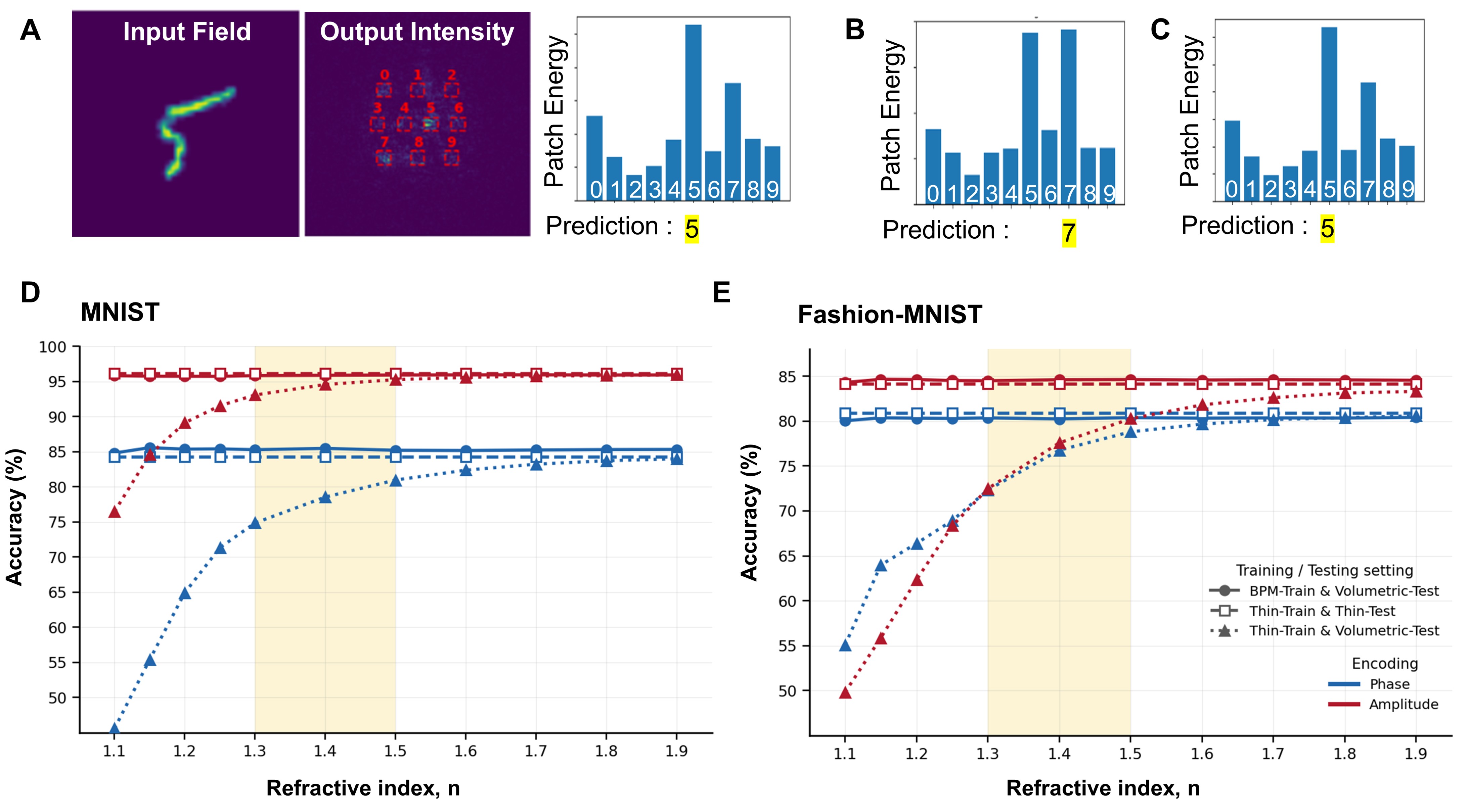}
    \caption{\textbf{Effect of model mismatch between thin-layer training and volumetric evaluation on optical classification.}
    (A) Representative MNIST input, detector output, and patch-energy distribution for a thin-trained/thin-tested model. 
    (B) The same thin-trained model evaluated with the finite-thickness volumetric forward model; the prediction becomes incorrect. 
    (C) A BPM-trained model evaluated with the same volumetric forward model; the correct class is recovered. 
    (D) MNIST classification accuracy for amplitude- and phase-encoded inputs. The yellow region highlights \(n\approx1.3\)--\(1.5\), a low/moderate-index range relevant to transparent visible-light-compatible polymeric, hydrogel, resin-based, silica-like, and scaffold-based platforms. 
    (E) Fashion-MNIST results show the same trend: the thin-trained mismatch is strongest in this highlighted regime, while BPM training better preserves volumetric-inference performance.}
    \label{fig:combined_images_fig2}
\end{figure}

 We utilized the MNIST and Fashion-MNIST datasets for the classification task. As shown in Fig.~\ref{fig:combined_images_fig2}(A), an input image was encoded as an optical field and propagated through a multi-layer \(D^2\)NN, producing an output intensity distribution at the detector plane. Classification was performed by directing optical energy into predefined detector regions, where each region corresponds to one class.

%We then examine how the forward-model assumption affects the detector-plane energy distribution. 
Fig.~\ref{fig:combined_images_fig2}(B--C) shows a representative example comparing thin-trained and BPM-trained models under volumetric evaluation. When the thin-trained model was evaluated using the same thin approximation used during training, it produced well-localized energy at the correct detector region. However, when the same thin-trained height profile was evaluated using the finite-thickness BPM forward model, the output intensity led to reduced detector contrast and incorrect energy allocation. In contrast, the BPM-trained model under the same volumetric evaluation correctly recovered the target class with high confidence (see the significantly stronger response at the correct class in Fig.~\ref{fig:combined_images_fig2}(C) ).

Fig.~\ref{fig:combined_images_fig2}(D--E) reports classification accuracy as a function of the refractive index for MNIST and Fashion-MNIST under amplitude and phase encoding. Under their native test configurations, both thin and $\partial$BPM-based models performed similarly well and reached the classification benchmarks previously demonstrated for \(D^2\)NNs~\cite{Lin2018Science, mengu2019analysis}. Note that our $\partial$BPM layers maintained the same learnability as the thin layers. Thin-trained models performed poorly when evaluated with the volumetric BPM model, suggesting a clear model mismatch. Interestingly, the model mismatch gradually diminished with increasing refractive index.  Transparent visible-light-compatible polymeric, hydrogel, resin-based, silica-like, and scaffold-based fabrication platforms occupy approximately \(n\approx1.3\)--\(1.5\) range (highlighted in yellow), while THz materials operate at refractive index close to 1.7~\cite{Lin2018Science,Rahman2023OcclusionCommunication,Mengu2023Multispectral}. Thus our results clearly show that thin-layer approximation disproportionately affects the visible range \(D^2\)NNs causing the models to fail.

%The degradation is strongest in the low- to moderate-index range highlighted in yellow, approximately \(n\approx1.3\)--\(1.5\), which is relevant to transparent visible-light-compatible polymeric, hydrogel, resin-based, silica-like, and scaffold-based fabrication platforms.

This trend follows from the fact that lower refractive index requires a larger physical relief thickness to achieve the same phase modulation. As a result, the layer deviates from the ideal thin-layer approximation and behaves more like a finite-thickness optical volume. In this regime, intra-layer diffraction, cumulative phase accumulation, and thickness-dependent field evolution become more significant. As the refractive index increases, the required relief height decreases, moving the system closer to the thin-layer limit and reducing the model mismatch between thin-layer and volumetric evaluation.

The larger gap observed for Fashion-MNIST shows the effect of model mismatch with increasing task complexity. Compared with MNIST, Fashion-MNIST contains more complex spatial structure and is therefore more sensitive to modeling errors that perturb the detector-plane energy distribution. Overall, these results show that finite-thickness modeling is especially important in the visible-compatible low- and moderate-index regime, where larger relative layer thickness (relative to the wavelength) amplifies the model mismatch between thin-layer training and volumetric evaluation.

\subsubsection{Results on the imaging tasks}

\begin{figure}[htbp]
    \centering
    \includegraphics[width=1\linewidth]{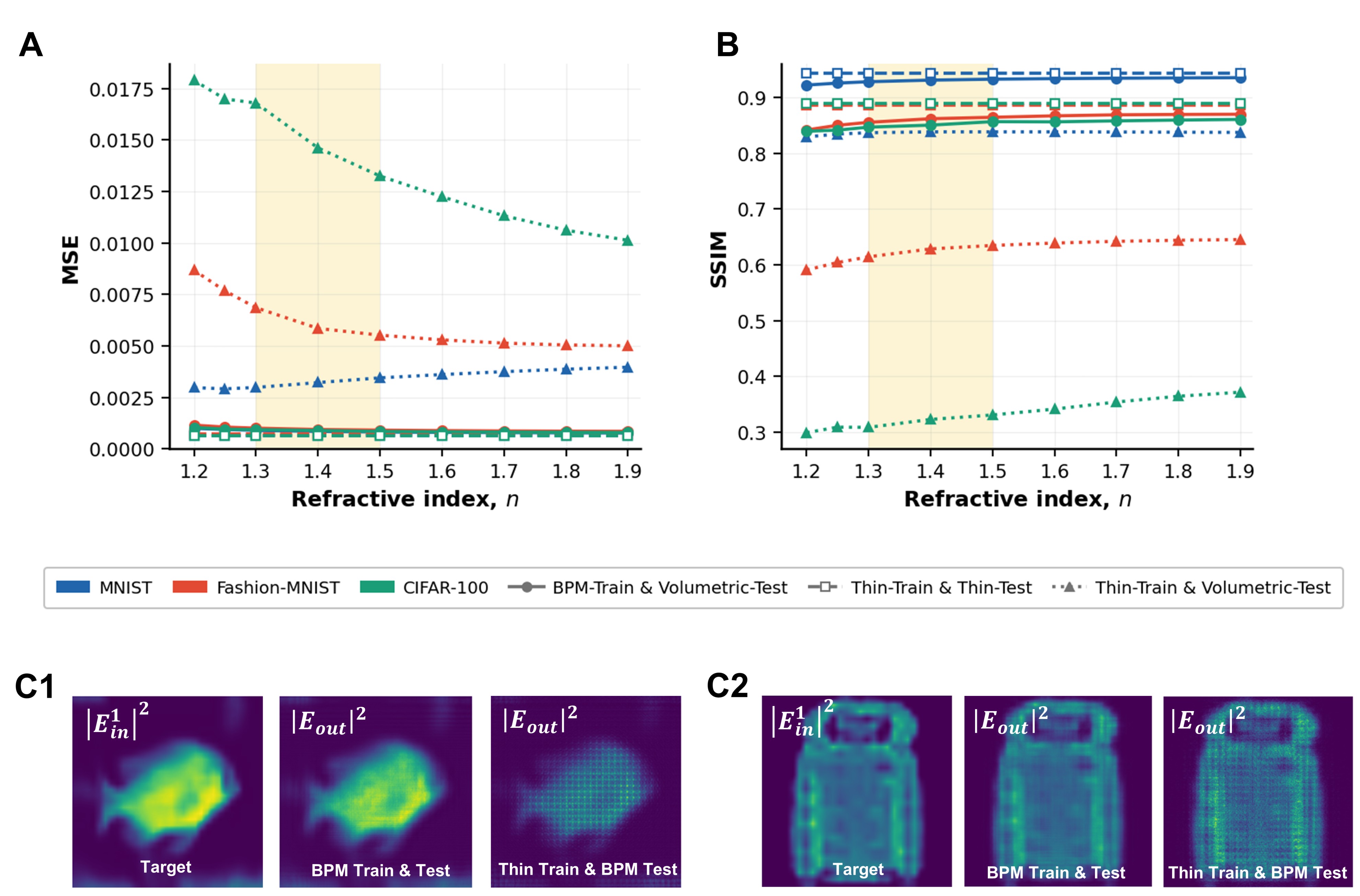}
    \caption{\textbf{Effect of model mismatch between thin-layer training and volumetric evaluation on optical imaging.}
    (A) MSE for MNIST, Fashion-MNIST, and CIFAR-100 imaging under thin-trained/thin-tested, thin-trained/volumetric-tested, and BPM-trained/volumetric-tested settings. 
    (B) Corresponding SSIM values. The yellow shaded region highlights \(n\approx1.3\)--\(1.5\), a low/moderate-index range relevant to transparent visible-light-compatible polymeric, hydrogel, resin-based, silica-like, and scaffold-based platforms. 
    (C1--C2) Representative reconstructions for CIFAR-100 and Fashion-MNIST at \(n=1.5\), showing the target, BPM-trained volumetric reconstruction, and thin-trained design evaluated with the volumetric forward model.}
    \label{fig:combined_images}
\end{figure}

Next, we evaluated the proposed framework on an imaging task with amplitude encoding. Unlike classification, where the final decision depends on the energy concentration within a small number of detector regions, imaging requires the full detector-plane intensity distribution to preserve spatial structure. For imaging we utilized the increasingly challenging datasets, MNIST, Fashion-MNIST, and CIFAR-100. 

Fig.~\ref{fig:combined_images}(A--B) reports the mean squared error (MSE) and Structural Similarity Index Measure (SSIM) as a function of refractive index. Thin-trained models performed well when evaluated under the same thin approximation used during training. As expected, models trained on the MNIST dataset performed best; models trained on Fashion-MNIST and CIFAR-100 performed similarly to each other but worse than MNIST.  Our $\partial$BPM layer based models performed slightly poorer than the thin layer models, especially for the Fashion-MNIST, and CIFAR-100 datasets. This performance drop suggests that there's room for improvement in learnability of $\partial$BPM layers. Finally, when the thin-trained height profiles were evaluated using the finite-thickness BPM forward model, reconstruction quality substantially degraded suggesting a significant model mismatch. Consistent with classification results, the model mismatch generally decreased with increasing refractive index. But even at THz materials' refractive index levels (i.e. closer to $n=1.7$) models still mismatched. As expected, the mismatch was smallest for the MNIST followed by Fashion-MNIST, and CIFAR-100. Fig.~\ref{fig:combined_images}(C1--C2) illustrate representative qualitative reconstructions at \(n=1.5\). Despite slightly worse SSIM levels, our $\partial$BPM layers  preserved the target structure in both CIFAR-100 and Fashion-MNIST examples.

%The degradation is most visible in the low- to moderate-index region highlighted in yellow, approximately \(n\approx1.3\)--\(1.5\), which corresponds to the material range relevant to many transparent visible-light-compatible volumetric fabrication platforms.

%In contrast, Differentiable BPM-$D^2$NN-trained models remained stable under volumetric evaluation. Across all datasets, designs trained and tested with the volumetric model maintained lower MSE and higher SSIM than thin-layer-trained designs evaluated with the volumetric model. This shows that the proposed training procedure learns fabrication-compatible height profiles whose outputs remain consistent when finite-thickness propagation is included.

%The representative reconstructions in Fig.~\ref{fig:combined_images}(C1--C2) illustrate this behavior at \(n=1.5\). For both CIFAR-100 and Fashion-MNIST, the BPM-trained model produces detector-plane reconstructions that better preserve the target structure. In contrast, the thin-trained design evaluated under the volumetric model shows stronger distortion and loss of structural detail. These differences are reflected most clearly in SSIM, indicating that structural fidelity is particularly sensitive to model mismatch between thin-layer training and volumetric evaluation.

Thus our results show a strong correlation between the dataset complexity and model mismatch (see Fig.~\ref{fig:combined_images}(B) ). MNIST remained comparatively robust because its images contain simpler spatial structure, while Fashion-MNIST and CIFAR-100 showed larger model mismatch, indicating imaging tasks with richer spatial details being more vulnerable to model mismatch. Overall, these results show that finite-thickness modeling is critical for reliable $D^2$NN imagers, especially in the visible-compatible low/moderate-index regime. 
%As refractive index decreases, larger normalized layer thickness increases intra-layer propagation effects, making the thin approximation less reliable. By training directly with the volumetric BPM forward model, the proposed framework preserves imaging performance under physically consistent evaluation and provides a wavelength-scalable route to more reliable diffractive optical reconstruction.

\subsection{FDTD full-wave validation of Differentiable BPM-$D^2$NN designs}
\label{subsec:fdtd_validation}

\begin{figure}[htbp]
    \centering
    \includegraphics[width=1\linewidth]{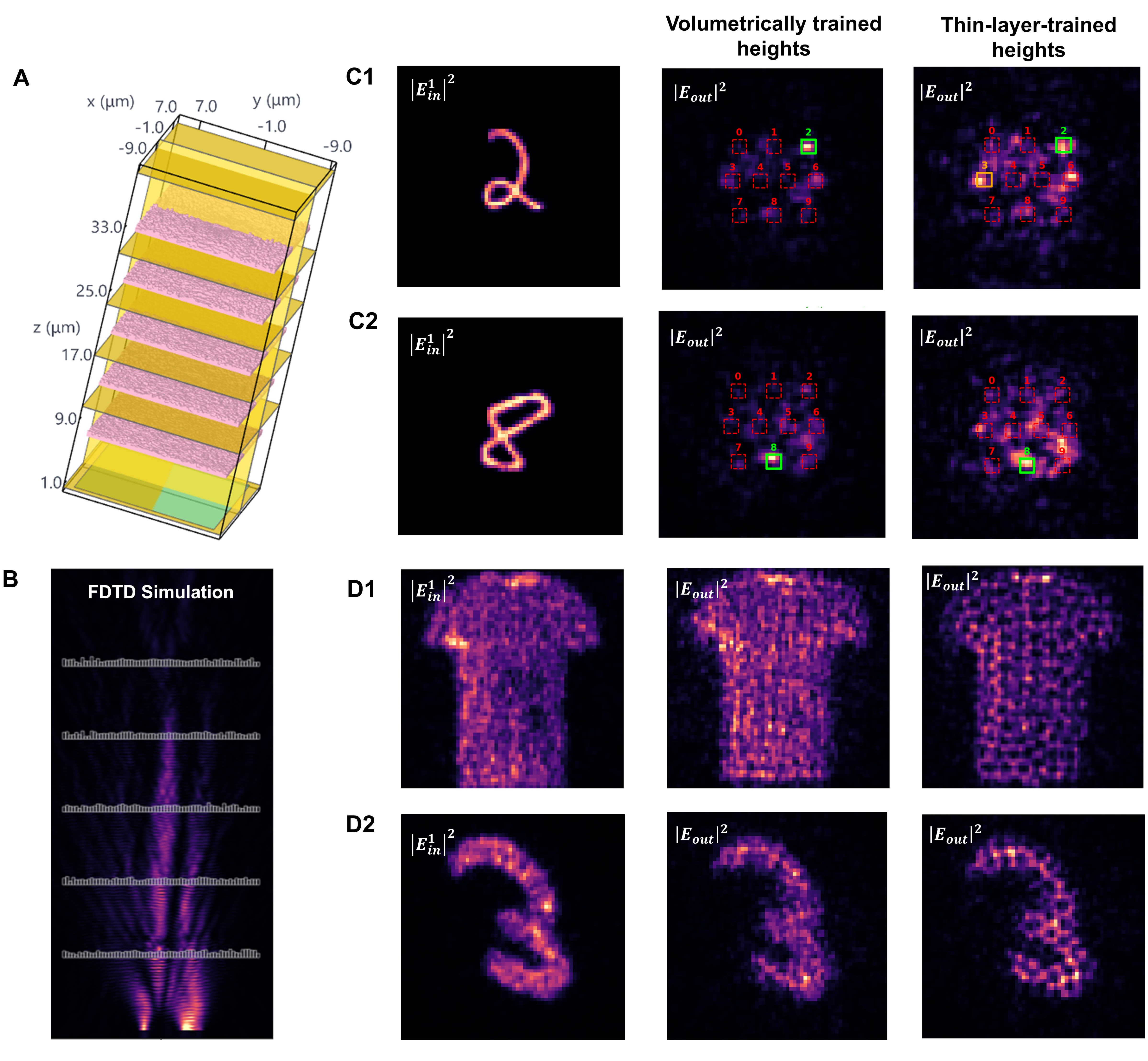}
\caption{\textbf{Full-wave FDTD validation of volumetrically trained and thin-layer-trained diffractive height maps.}
(A) Three-dimensional rendering of the five-layer finite-thickness diffractive structure used in FDTD simulation.
(B) Cross-sectional FDTD field intensity along the propagation axis, showing intra-layer and inter-layer field evolution.
(C1--C2) Classification examples evaluated directly in FDTD. The left column shows the input intensity $|E_{\mathrm{in}}|^2$, and the middle and right columns show $|E_{\mathrm{out}}|^2$ from volumetrically trained and thin-layer-trained height maps, respectively. Red dashed boxes indicate class-detector regions, and the green box marks the target detector. Volumetrically trained height maps show stronger target-region localization.
(D1--D2) Imaging examples evaluated directly in FDTD. The left column shows the input/target intensity $|E_{\mathrm{in}}|^2$, while the middle and right columns show reconstructed outputs $|E_{\mathrm{out}}|^2$ from volumetrically trained and thin-layer-trained height maps. Volumetrically trained height maps better preserve the target spatial structure under full-wave evaluation.}

    \label{fig:combined_images_fig4}
\end{figure}

We used full-wave finite-difference time-domain (FDTD) simulations as an independent physical reference to evaluate the learned diffractive structures. FDTD solves Maxwell's equations on a fine electromagnetic mesh and was not used for optimization, and thereby serves as a post-training validation step much closer to the physics than the scalar theory used during training. We exported the trained height maps from the learning model, converted them into three-dimensional dielectric structures, and evaluated without any re-optimization (see Fig.~\ref{fig:combined_images_fig4}(A--B)).

This validation addresses the practical question of whether the proposed volumetric training procedure produces height maps that remain functional under a higher-fidelity full-wave model. We therefore compared two types of learned structures under the same FDTD setting: height maps trained with the thin-layers and height maps trained with $\partial$BPM layers. The FDTD model used the same wavelength, refractive index, aperture, layer spacing, input fields, detector regions, and exported height profiles as the corresponding optical design. Because FDTD is computationally expensive, the validation was performed on randomly selected subsets of ten test samples for classification and imaging at \(n=1.5\). FDTD detector-plane intensities were area-averaged onto the \(64\times64\) detector grid before metric computation, and a single global intensity scale factor was fitted before error metrics were calculated.

%Fig.~\ref{fig:combined_images_fig4} provides qualitative full-wave evidence supporting the quantitative FDTD results. The rendered structure and cross-sectional field map show that the exported finite-thickness layers are evaluated as three-dimensional dielectric volumes, allowing FDTD to capture both intra-layer and inter-layer field evolution. In the classification examples, the Differentiable BPM-$D^2$NN-trained height maps route more energy into the target detector region, while the thin-layer-trained height maps produce more dispersed detector-plane energy. In the imaging examples, the volumetrically trained height maps better preserve the target spatial structure, whereas the thin-layer-trained height maps show stronger degradation under the same full-wave evaluation.

\paragraph{Classification validation.}
For classification, the relevant physical output is the energy routed into the predefined class-detector regions. We therefore evaluated the FDTD output using task-level detector metrics. Accuracy measures the final optical decision. The correct-patch fraction \(p_y\) measures the fraction of detector-patch energy assigned to the true class. The margin ratio measures how strongly the correct detector patch exceeds the strongest competing patch. The correct efficiency \(\eta_{\mathrm{correct}}\) measures how much of the total detector-plane energy reaches the true class region. Cross-entropy measures the confidence of the normalized detector-patch distribution with respect to the true label.

\begin{table}[htbp]
\centering
\scriptsize
\setlength{\tabcolsep}{4pt} 
\caption{\textbf{Full-wave FDTD classification performance for thin-layer-trained and Differentiable BPM-$D^2$NN-trained height maps.} 
Both designs were evaluated directly in FDTD over the same ten selected test samples. Patch energies were integrated over identical physical detector regions.}
\label{tab:fdtd_classification_performance}
\rowcolors{2}{gray!7}{white}
\begin{tabularx}{\columnwidth}{>{\raggedright\arraybackslash}Xccccc}
\toprule
\rowcolor{gray!18}
\textbf{Height map} & \textbf{Acc. $\uparrow$} & \textbf{$p_y$ $\uparrow$} & \textbf{\shortstack{Margin\\ratio $\uparrow$}} & \textbf{$\eta_{\mathrm{correct}}$ $\uparrow$} & \textbf{CE $\downarrow$} \\
\midrule
Thin-layer-trained heights & 50\% & $0.2090 \pm 0.0568$ & $1.1704 \pm 0.4621$ & $0.0015 \pm 0.0006$ & $1.6006 \pm 0.2637$ \\
Differentiable BPM-$D^2$NN-trained heights & 90\% & $0.2516 \pm 0.0450$ & $1.4834 \pm 0.4088$ & $0.0022 \pm 0.0006$ & $1.3965 \pm 0.1853$ \\
\bottomrule
\end{tabularx}
\end{table}

Table~\ref{tab:fdtd_classification_performance} shows that Differentiable BPM-$D^2$NN-trained height maps produce stronger classification behavior under full-wave FDTD evaluation. Compared with thin-layer-trained height maps, volumetrically trained structures improved FDTD accuracy from \(50\%\) to \(90\%\), while also improving all the other metrics. Fig.~\ref{fig:combined_images_fig4}(C1--C2) shows two representative examples. 

%increased the correct-patch energy fraction, increased the margin ratio, improved correct efficiency, and reduced cross-entropy. These results indicate that volumetric training reduces the model mismatch that appears when thin-layer-trained designs are evaluated as finite-thickness physical structures.

\paragraph{Imaging validation.}
For imaging, the relevant physical output is the full detector-plane intensity distribution. We evaluated the FDTD output against the target image using Pearson correlation, scale-invariant relative \(L_2\) error, and normalized root mean squared error (NRMSE). Pearson correlation measures spatial agreement with the target image, while relative \(L_2\) error and NRMSE quantify normalized reconstruction error after global intensity scaling. As shown in Table~\ref{tab:fdtd_imaging_performance} $D^2$NN-designs using $\partial$BPM layers better preserve the target image under full-wave propagation on all metrics. Fig.~\ref{fig:combined_images_fig4}(D1--D2) shows two representative examples. 

\begin{table}[htbp]
\centering
\scriptsize
\caption{\textbf{Full-wave FDTD imaging performance for thin-layer-trained and Differentiable BPM-$D^2$NN-trained height maps.}
Metrics compare the FDTD detector-plane output with the corresponding target image over the same ten selected test samples. FDTD outputs were evaluated after grid matching and global intensity scaling.}
\label{tab:fdtd_imaging_performance}
\rowcolors{2}{gray!7}{white}
\begin{tabular}{lccc}
\toprule
\rowcolor{gray!18}
\textbf{Height map} &
\textbf{Pearson \(r\) $\uparrow$} &
\textbf{Rel. \(L_2\) $\downarrow$} &
\textbf{NRMSE $\downarrow$} \\
\midrule
Thin-layer-trained heights &
\(0.7838 \pm 0.0167\) &
\(0.5540 \pm 0.0162\) &
\(0.1073 \pm 0.0068\) \\
Differentiable BPM-$D^2$NN-trained heights &
\(0.8168 \pm 0.0095\) &
\(0.5168 \pm 0.0136\) &
\(0.0935 \pm 0.0070\) \\
\bottomrule
\end{tabular}
\end{table}

%Table~\ref{tab:fdtd_imaging_performance} shows that Differentiable BPM-$D^2$NN-trained height maps better preserve the target image under full-wave propagation. Compared with thin-layer-trained height maps, volumetrically trained designs improved the average output--target Pearson correlation from \(0.7838\) to \(0.8168\), reduced the relative \(L_2\) error from \(0.5540\) to \(0.5168\), and reduced NRMSE from \(0.1073\) to \(0.0935\). These improvements show that the proposed volumetric training procedure improves physical reliability not only for detector-based classification but also for spatial image formation.

Overall, our Differentiable BPM layer provides a practical intermediate model between thin-layer training and direct full-wave electromagnetic optimization. Although full-wave Maxwell solvers such as FDTD provide higher-fidelity electromagnetic modeling, using them directly inside end-to-end $D^2$NN training would be computationally prohibitive because each optimization step would require large three-dimensional subwavelength simulations and corresponding gradient computations over a large 3D grid. In contrast, our $\partial$BPM layers are computationally tractable for efficient training while minimizing the model mismatch due to intra-layer propagation. FDTD can then be used as an independent physical reference to test whether the trained structures preserve their intended optical behavior under full-wave Maxwell equations. 

%The FDTD results therefore support the role of Differentiable BPM-$D^2$NN as a trainable physics-aware bridge between fast thin-layer models and expensive full-wave validation.

%Differentiable BPM-$D^2$NN is therefore used here as a tractable volumetric forward model: it captures the dominant finite-thickness intra-layer propagation effects ignored by the thin-layer approximation while remaining efficient enough for iterative learning. Full-wave FDTD is then used as an independent physical reference to test whether the trained structures preserve their intended optical behavior under Maxwell-equation-based propagation. The FDTD results therefore support the role of Differentiable BPM-$D^2$NN as a trainable physics-aware bridge between fast thin-layer models and expensive full-wave validation.

\section{Summary}

%This allows the network to account for intra-layer diffraction, cumulative phase accumulation, and the thickness-dependent reduction of free-space distance between adjacent layers.

We introduced the Differentiable BPM ($\partial$BPM) layer for diffractive deep neural networks ($D^2$NNs) to eliminate the model mismatch due to the conventional thin-layer approximation. Our $\partial$BPM layer discretizes a fabrication-compatible volumetric refractive-index structure into axial sub-slices, and propagates the optical field through the material. A differentiable soft occupancy function makes our $\partial$BPM layers end-to-end differentiable. Across optical classification and imaging tasks, the proposed framework consistently reduced the model mismatch, especially at low and moderate refractive indices, where larger physical thickness is required to achieve the same phase modulation. This regime is important for visible-light-compatible polymeric, hydrogel, resin-based, silica-like, and scaffold-based fabrication platforms, which typically use low- to moderate-index materials.

Full-wave FDTD simulations further supported the physical reliability of the proposed approach. We used FDTD as an independent physical reference to evaluate learned height maps after training, without re-optimization. When thin-layer-trained and $\partial$BPM-layer-trained height maps were both evaluated under FDTD, the volumetrically trained structures showed more reliable physical behavior, including higher classification accuracy, stronger correct-patch energy concentration, and lower normalized imaging error. Overall, Differentiable BPM-$D^2$NN provides a physics-aware bridge between fast thin-layer training and expensive full-wave electromagnetic validation. By embedding finite-thickness propagation directly into the training loop, the proposed framework produces diffractive neural networks that are more reliable, fabrication-compatible, and wavelength-scalable. This makes volumetric training an important step toward practical visible-light and low/moderate-index free-space optical neural networks.

%These results indicate that Differentiable BPM-$D^2$NN captures the dominant finite-thickness effects needed for practical diffractive design while remaining more tractable than direct full-wave electromagnetic optimization.

\section{Method Details}
\label{sec:method}

\subsection{Task-specific configurations}

The optical simulations are formulated in wavelength-normalized units. In this representation, the behavior of the diffractive system is governed primarily by dimensionless ratios such as aperture-to-wavelength, pixel-pitch-to-wavelength, layer-spacing-to-wavelength, and thickness-to-wavelength. Therefore, the same normalized optical design can be instantiated at different physical wavelengths by scaling the transverse and axial dimensions proportionally with \(\lambda\). This wavelength-normalized formulation follows the geometric scaling principle used in free-space diffractive optical neural networks, including the original 3D-printed D\(^{2}\)NN demonstration~\cite{Lin2018Science}.

For the main task-level experiments, we used a visible-light wavelength of \(\lambda=532~\mathrm{nm}\). This choice emphasizes the visible-light regime, where transparent low- and moderate-index materials make finite-thickness effects especially relevant. Although the absolute wavelength is changed, the optical geometry is defined through normalized ratios, so the same physical trends are expected to hold when the geometry is scaled consistently with \(\lambda\). To verify that the observed model mismatch between thin-layer training and volumetric evaluation is not specific to the visible wavelength scale, we also repeated the experiments at \(\lambda=750~\mu\mathrm{m}\), matching the terahertz-scale wavelength used in the original 3D-printed free-space D\(^{2}\)NN demonstration~\cite{Lin2018Science}. These wavelength-scaled THz results are reported in the Appendix.

The FDTD validation experiments were performed at \(\lambda=532~\mathrm{nm}\) using a matched \(64\times64\) geometry. This reduced resolution was chosen to make full-wave electromagnetic simulation computationally tractable. The comparison was not made between a high-resolution training model and a reduced FDTD model; instead, the thin-layer-trained and Differentiable BPM-\(D^2\)NN-trained designs were evaluated in FDTD using the same \(64\times64\) physical geometry, including the same wavelength-scaled aperture, layer spacing, refractive index, input field, detector regions, and exported finite-thickness height profiles.

\subsubsection{Classification}

For classification experiments, we used a 5-layer diffractive network with geometry defined relative to the wavelength. Specifically, the inter-layer spacing was \(40\lambda\), the aperture was approximately \(107\lambda\), the pixel pitch was approximately \(0.53\lambda\), and the detector patch width was \(6.4\lambda\). In the main visible-light instantiation at \(\lambda=532~\mathrm{nm}\), all transverse and axial dimensions were obtained by multiplying these normalized values by the wavelength. The same normalized geometry was also instantiated at \(\lambda=750~\mu\mathrm{m}\) for the wavelength-scaled THz comparison reported in the Appendix. Input images were resized to \(112\times112\) and zero-padded to the simulation grid.

Two input encodings were considered.

\paragraph{Phase encoding.}
\begin{equation}
U_0(x,y)=\exp\!\big(i\,2\pi X(x,y)\big),
\end{equation}

\paragraph{Amplitude encoding.}
\begin{equation}
U_0(x,y)=X(x,y),
\end{equation}
where \(X(x,y)\in[0,1]\) is the normalized image.

The output plane was partitioned into 10 detector regions, one per class. The detector response for class \(c\) was
\begin{equation}
p_c=\sum_{(x,y)\in\Omega_c} I_{\text{out}}(x,y).
\end{equation}
A scaled-logit form was used:
\begin{equation}
\tilde p_c = 10\,\frac{p_c}{\max_j p_j+\epsilon},
\end{equation}
and the classification loss was
\begin{equation}
\mathcal{L}_{\text{cls}}
=
-\log\frac{\exp(\tilde p_y)}{\sum_{c=1}^{10}\exp(\tilde p_c)}.
\end{equation}

\subsubsection{Imaging}

For imaging experiments, we used a 5-layer D\(^{2}\)NN with a normalized optical geometry, consistent with the design principles introduced by Lin et al.~\cite{Lin2018Science}. The inter-layer spacing was set to approximately \(5.33\lambda\), while the propagation distance from the final layer to the output plane was approximately \(9.33\lambda\). The system aperture was \(120\lambda\), and the pixel pitch was \(0.4\lambda\). The main imaging experiments were instantiated at \(\lambda=532~\mathrm{nm}\), and the same wavelength-normalized configuration was also evaluated at \(\lambda=750~\mu\mathrm{m}\) in the Appendix to confirm that the observed behavior follows the normalized optical scaling.

For each dataset, \(N_{\text{train}}=1500\), \(N_{\text{val}}=200\), and \(N_{\text{test}}=300\) samples were randomly selected from the full dataset.

The input field was amplitude-only and defined as
\begin{equation}
U_0(x,y)=A(x,y),
\end{equation}
where \(A(x,y)\in[0,1]\) denotes the normalized grayscale image. The diffractive network applies a linear transformation to the input optical field, which can be written as
\begin{equation}
U_{\text{out}}(x,y)=\mathcal{H}\big(U_0(x,y)\big),
\end{equation}
where \(\mathcal{H}(\cdot)\) denotes the overall linear propagation and phase modulation operator implemented by the D\(^{2}\)NN.

At the detector plane, only the intensity of the optical field is measured:
\begin{equation}
I_{\text{out}}(x,y)=\left|U_{\text{out}}(x,y)\right|^2
=\left|\mathcal{H}\big(U_0(x,y)\big)\right|^2.
\end{equation}
Thus, the D\(^{2}\)NN performs a linear transformation in the field domain followed by a nonlinear intensity measurement at the detector.

The target output was defined as
\begin{equation}
T(x,y)=A^2(x,y),
\end{equation}
which corresponds to the desired intensity distribution associated with the input amplitude.

The network was trained to regress the measured output intensity toward this target using a mean squared error (MSE) loss:
\begin{equation}
\mathcal{L}_{\text{img}}
=
\frac{1}{HW}\sum_{x,y}\left(I_{\text{out}}(x,y)-T(x,y)\right)^2.
\end{equation}

\subsubsection{FDTD simulations}
\label{subsubsec:fdtd_methods}

We used full-wave finite-difference time-domain (FDTD) simulations as an independent electromagnetic reference for evaluating learned diffractive structures after training. FDTD simulations were performed using the Tidy3D solver (Flexcompute Inc.)~\cite{flexcompute_tidy3d}. The purpose of these simulations was not to replace Differentiable BPM-\(D^2\)NN during training, but to test whether the trained height maps preserved their intended optical behavior under a Maxwell-equation-based solver.

For each selected sample, the learned diffractive layers were exported as physical height maps and converted into three-dimensional dielectric structures. The FDTD model used the same wavelength, refractive index, aperture, layer spacing, detector regions, and height profiles as the corresponding design. Unless otherwise stated, the full-wave validation was performed at \(\lambda=532~\mathrm{nm}\), corresponding to the visible-light-scaled setting used in the main experiments. The FDTD mesh used a minimum spatial resolution of six grid points per wavelength, with additional refinement in the diffractive-layer region when required. The same input optical field used in the learning model was injected into the FDTD domain using a custom field source. To avoid source-plane artifacts, field comparisons near the input were performed at a small propagation distance after the source plane rather than exactly on the source sheet.

Because the FDTD solver represents the field on a finer electromagnetic mesh than the \(64\times64\) detector grid, detector-plane comparisons require a common spatial grid. The native FDTD detector intensity was area-averaged onto the detector grid before metric computation. For a detector pixel region \(\Omega_{ij}\), the corresponding binned FDTD intensity was computed as
\begin{equation}
I^{\mathrm{bin}}_{\mathrm{FDTD}}(i,j)
=
\frac{1}{|\Omega_{ij}|}
\int_{\Omega_{ij}}
I_{\mathrm{FDTD}}(x,y)\,dx\,dy .
\end{equation}
This produces an FDTD intensity map with the same spatial sampling as the detector output used for evaluation. Since FDTD outputs can differ by an arbitrary global intensity scale due to source normalization, solver conventions, and monitor definitions, we fit a single global scaling factor before computing image-level error metrics:
\begin{equation}
\alpha
=
\frac{
\left\langle I_{\mathrm{FDTD}}, I_{\mathrm{ref}}\right\rangle
}{
\left\langle I_{\mathrm{FDTD}}, I_{\mathrm{FDTD}}\right\rangle
}.
\end{equation}
Here, \(I_{\mathrm{FDTD}}\) denotes the FDTD intensity after grid matching, and \(I_{\mathrm{ref}}\) denotes the corresponding reference intensity on the same grid. For imaging performance, \(I_{\mathrm{ref}}\) is the target image intensity.

For image-level FDTD evaluation, we report Pearson correlation, scale-invariant relative \(L_2\) error, and normalized root mean squared error (NRMSE). Pearson correlation measures spatial pattern agreement independent of global brightness. The scale-invariant relative \(L_2\) error is defined as
\begin{equation}
\mathrm{Rel.}\ L_2
=
\frac{
\left\|
\alpha I_{\mathrm{FDTD}} - I_{\mathrm{ref}}
\right\|_2
}{
\left\|
I_{\mathrm{ref}}
\right\|_2
}.
\end{equation}
The RMSE after global scaling is
\begin{equation}
\mathrm{RMSE}
=
\sqrt{
\frac{1}{N}
\sum_i
\left(
\alpha I_{\mathrm{FDTD},i}
-
I_{\mathrm{ref},i}
\right)^2
},
\end{equation}
and the normalized RMSE is computed as
\begin{equation}
\mathrm{NRMSE}
=
\frac{\mathrm{RMSE}}
{I_{\mathrm{ref}}^{\max}-I_{\mathrm{ref}}^{\min}+\epsilon},
\end{equation}
where \(N\) is the number of pixels and \(\epsilon\) is a small constant for numerical stability.

For classification, the final decision is determined by the optical energy integrated inside predefined detector patches. Let \(\Omega_k\) denote the physical detector region corresponding to class \(k\). The class-patch energy is
\begin{equation}
E_k
=
\int_{\Omega_k}
I(x,y)\,dx\,dy ,
\end{equation}
and the predicted class is
\begin{equation}
\hat{y}
=
\arg\max_k E_k .
\end{equation}

To evaluate physical classification performance under FDTD, both thin-layer-trained and Differentiable BPM-\(D^2\)NN-trained height maps were simulated directly in FDTD using the same selected test samples. For a sample with true class \(y\), the correct-patch fraction is
\begin{equation}
p_y
=
\frac{E_y}{\sum_k E_k}.
\end{equation}
The margin ratio is defined as
\begin{equation}
\mathrm{Margin\ Ratio}
=
\frac{
E_y
}{
\max_{k\neq y}E_k+\epsilon
},
\end{equation}
where values larger than one indicate that the correct detector patch receives more energy than any competing class patch. We also compute the correct efficiency,
\begin{equation}
\eta_{\mathrm{correct}}
=
\frac{E_y}{E_{\mathrm{total}}},
\end{equation}
where \(E_{\mathrm{total}}\) is the total detector-plane energy. Finally, we compute cross-entropy from the normalized patch-energy distribution:
\begin{equation}
\mathrm{CE}
=
-\log(p_y+\epsilon).
\end{equation}
These metrics jointly measure the final prediction, the concentration of energy in the correct detector region, the separation between the correct and strongest competing detector regions, and the confidence of the physical optical decision.

For imaging, the FDTD output was evaluated as a full detector-plane intensity distribution rather than as a set of class-patch energies. Thin-layer-trained and Differentiable BPM-\(D^2\)NN-trained height maps were simulated directly in FDTD and compared with the target image after grid matching and global intensity scaling. In the identity-imaging setting used here, the target is the input intensity pattern. We report Pearson correlation, scale-invariant relative \(L_2\) error, and NRMSE. These metrics directly quantify spatial agreement and normalized reconstruction error while avoiding dependence on arbitrary absolute intensity scaling.

For both classification and imaging, FDTD validation was performed on randomly selected subsets of ten test samples. This subset size was chosen because each full-wave simulation requires a three-dimensional electromagnetic solve at subwavelength spatial resolution for a multi-layer dielectric structure, making exhaustive FDTD evaluation computationally impractical. The FDTD study is therefore used as a targeted physical validation rather than as the primary statistical evaluation; the main task-level trends are quantified using the differentiable volumetric forward model over full test sets and refractive-index sweeps. To strengthen the comparison despite the small FDTD subset, the same selected samples were used when comparing thin-layer-trained and Differentiable BPM-\(D^2\)NN-trained height maps, enabling a paired comparison under identical FDTD conditions. All FDTD simulations used the exported height maps without re-optimization, so the reported results measure how well each learned structure transfers directly to a full-wave electromagnetic solver.

\subsection{Training and evaluation protocol}

All models were optimized end-to-end with Adam. For Differentiable BPM-\(D^2\)NN training, each diffractive layer was discretized into \(N_{\text{sub}}=15\) axial sub-slices. The thin-layer baseline applied the learned phase modulation at a single axial plane without intra-layer propagation. The sigmoid temperature \(\tau\) was annealed during volumetric training to gradually sharpen the slice boundaries.

For classification, Adam was used with learning rate \(10^{-3}\), batch size \(128\), and \(10\) epochs. For imaging, Adam was used with learning rate \(10^{-3}\), batch size \(4\), and \(50\) epochs. MNIST and FashionMNIST were resized to \(112\times112\) and padded to the classification grid \((200\times200)\), while CIFAR-100 was converted to grayscale and resized to the imaging grid \((300\times300)\). MNIST and FashionMNIST were also mapped to the imaging grid for the corresponding imaging experiments.

For each refractive index in the material sweep, we evaluated three settings:
\begin{enumerate}
\item \textbf{Differentiable BPM-\(D^2\)NN train \(\rightarrow\) volumetric test:} volumetric training and volumetric evaluation,
\item \textbf{Thin-layer train \(\rightarrow\) thin-layer test:} thin-layer training and thin-layer evaluation,
\item \textbf{Thin-layer train \(\rightarrow\) volumetric test:} thin-layer training followed by evaluation under the volumetric forward model.
\end{enumerate}
The third setting quantifies the model mismatch introduced when a design trained under the thin-layer approximation is evaluated as a finite-thickness structure.

For the FDTD validation experiments, all designs were evaluated using the same \(64\times64\) diffractive-layer grid. This reduced grid was used to make full-wave electromagnetic simulation computationally feasible while keeping the optical geometries matched across training models. The Differentiable BPM-\(D^2\)NN framework itself is not restricted to \(64\times64\) grids; larger grids were used in the main task-level experiments, while the \(64\times64\) setting was used specifically for full-wave FDTD validation.

For the classification FDTD validation, we used a wavelength of \(\lambda=532~\mathrm{nm}\), pixel pitch \(\Delta x=266~\mathrm{nm}\), aperture \(64\Delta x=17.024~\mu\mathrm{m}\), five diffractive layers, and inter-layer spacing \(z=6.4~\mu\mathrm{m}\). For the imaging FDTD validation, we used the same wavelength, pixel pitch, aperture, and number of layers, with inter-layer spacing \(z=1.25~\mu\mathrm{m}\) and final propagation distance \(z_{\mathrm{out}}=1.50~\mu\mathrm{m}\). In these validation experiments, thin-layer-trained and Differentiable BPM-\(D^2\)NN-trained models were trained using the same optimizer, learning rate, batch size, and number of epochs as their corresponding task settings described above. Therefore, the observed differences arise from the forward optical model rather than from differences in training resolution, optimization settings, or numerical hyperparameters.

%%%%%%%%%%%%%%%%%%%%%%% References %%%%%%%%%%%%%%%%%%%%%%%%%

%%%%%%%%%% If using BibTeX:
\bibliography{sample}
\newpage

\title{Beyond the Thin-Layer Limit: Differentiable Volumetric Training for Visible-Range Diffractive Neural Networks - Supplementary Material}

\setcounter{section}{0} % Resets the section counter to 0

\section{Validation at THz Wavelengths}
\label{app:thz_validation}

The main experiments in this work are reported at the visible wavelength of \(\lambda=532~\mathrm{nm}\), where low- and moderate-index transparent materials make finite-thickness effects especially relevant. To verify that the observed thin-to-volumetric mismatch is not specific to this absolute wavelength, we repeat the same BPM experiments under a wavelength-scaled THz configuration. Specifically, we instantiate the normalized optical geometry at \(\lambda=750~\mu\mathrm{m}\), matching the wavelength used in the original 3D-printed free-space D\(^{2}\)NN demonstration. All relevant physical dimensions, including aperture, pixel pitch, layer spacing, detector layout, and layer thickness, are scaled consistently with \(\lambda\).

\begin{figure}[htbp]
    \centering
    \includegraphics[width=1\linewidth]{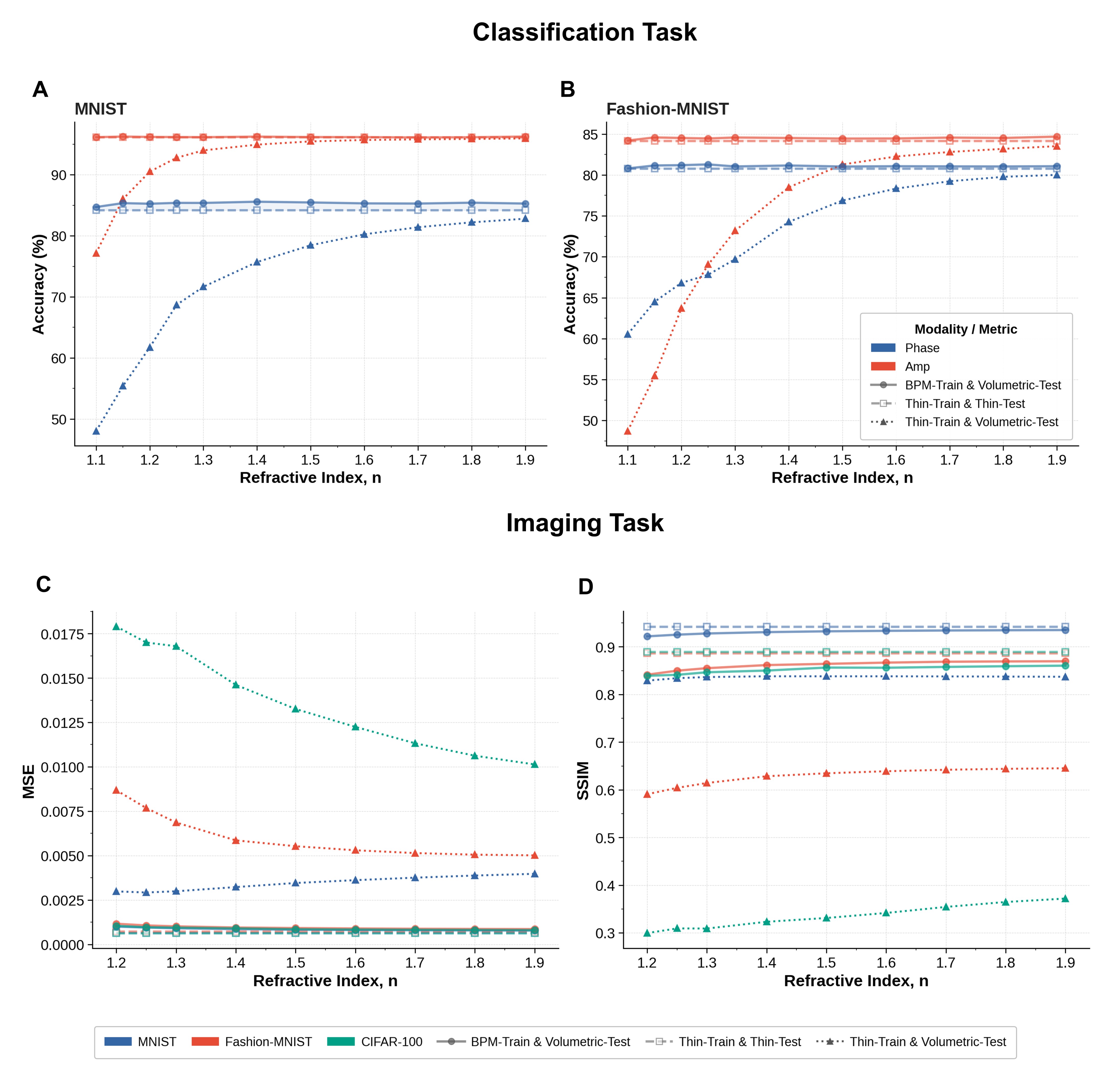}
    \caption{\textbf{Wavelength-scaled THz validation of the thin-to-volumetric model mismatch.}
    The same normalized optical geometries used in the main experiments are instantiated at \(\lambda=750~\mu\mathrm{m}\), matching the THz wavelength used in the original free-space D\(^{2}\)NN demonstration. 
    (A--B) Classification accuracy for MNIST and Fashion-MNIST under phase and amplitude encoding. Thin-trained models degrade when evaluated with the volumetric BPM forward model, especially at lower refractive indices, while BPM-trained models preserve stable volumetric-test performance. 
    (C--D) Imaging performance across MNIST, Fashion-MNIST, and CIFAR-100, reported using MSE and SSIM. The thin-trained/volumetric-tested setting shows the largest reconstruction degradation at low refractive index, while BPM-trained/volumetric-tested models remain stable. These THz-scaled results confirm that the observed mismatch is controlled by wavelength-normalized geometry and persists across wavelength regimes.}
    \label{fig:bpm_fig5_thz}
\end{figure}

Fig.~\ref{fig:bpm_fig5_thz} shows that the same qualitative behavior is preserved in the THz-scaled setting. For classification, thin-trained models perform well when evaluated under the same thin approximation used during training, but degrade when the corresponding height profiles are evaluated with the volumetric BPM forward model. This degradation is strongest at lower refractive indices and decreases as \(n\) increases. In contrast, BPM-trained models maintain stable performance under volumetric evaluation for both MNIST and Fashion-MNIST.

The imaging results show the same trend. Thin-trained designs evaluated with the volumetric model exhibit higher MSE and lower SSIM, especially at lower refractive indices. BPM-trained designs remain stable and closely follow the matched volumetric behavior across MNIST, Fashion-MNIST, and CIFAR-100.

\section{FDTD Simulation Details}
\label{app:fdtd_details}

\subsection{FDTD source construction and numerical discretization}
\label{app:fdtd_source_grid}

For the BPM--FDTD validation experiments, the learned diffractive layers are exported as physical height maps and reconstructed as finite-thickness dielectric structures in the FDTD solver. The validation uses a matched \(64\times64\) diffractive-layer grid to keep the full-wave simulations computationally feasible while preserving the same physical aperture, pixel pitch, layer spacing, wavelength, and refractive index used in the corresponding BPM model. The BPM framework itself is not restricted to this resolution.

The input field exported from BPM is the complex scalar field at the input plane,
\begin{equation}
u_0(x,y)=u_{\mathrm{real}}(x,y)+i u_{\mathrm{imag}}(x,y).
\end{equation}
For amplitude-encoded inputs, \(u_0(x,y)\) corresponds to the normalized image amplitude. Since FDTD solves Maxwell's equations rather than a scalar diffraction equation, this scalar BPM field must be converted into a vector electromagnetic source. We use a forward-propagating angular-spectrum construction so that the injected source satisfies the transverse Maxwell condition for each propagating spatial-frequency component.

The BPM field is first assigned to the dominant \(y\)-polarized electric-field component,
\begin{equation}
E_y(x,y,z_0)=u_0(x,y),
\end{equation}
and transformed into the spatial-frequency domain:
\begin{equation}
E_y(f_x,f_y)=\mathcal{F}\{u_0(x,y)\}.
\end{equation}
For each spectral component, we define the normalized transverse wave-vector components
\begin{equation}
\alpha=\lambda f_x,
\qquad
\beta=\lambda f_y,
\end{equation}
and the forward longitudinal component
\begin{equation}
\gamma=\sqrt{1-\alpha^2-\beta^2}.
\end{equation}
Only components satisfying
\begin{equation}
1-\alpha^2-\beta^2>0
\end{equation}
are retained. Evanescent or grazing components are removed to form a stable one-way forward source. 

To construct a Maxwell-consistent electric field, we set
\begin{equation}
E_x(f_x,f_y)=0
\end{equation}
and enforce the transversality condition
\begin{equation}
\bm{k}\cdot\bm{E}=0.
\end{equation}
Using the normalized wave-vector components, this gives
\begin{equation}
\alpha E_x+\beta E_y+\gamma E_z=0.
\end{equation}
Since \(E_x=0\), the longitudinal electric-field component is
\begin{equation}
E_z(f_x,f_y)=-\frac{\beta}{\gamma}E_y(f_x,f_y).
\end{equation}
Thus, the electric field spectrum used for the FDTD source is
\begin{equation}
\bm{E}(f_x,f_y)=
\begin{bmatrix}
0\\[3pt]
E_y(f_x,f_y)\\[3pt]
-\dfrac{\beta}{\gamma}E_y(f_x,f_y)
\end{bmatrix}.
\end{equation}

The magnetic field is then computed from the plane-wave Maxwell relation
\begin{equation}
\bm{H}(f_x,f_y)=\frac{1}{\eta_0}\hat{\bm{k}}(f_x,f_y)\times \bm{E}(f_x,f_y),
\end{equation}
where \(\eta_0\) is the free-space impedance and
\begin{equation}
\hat{\bm{k}}(f_x,f_y)=
\begin{bmatrix}
\alpha\\
\beta\\
\gamma
\end{bmatrix}.
\end{equation}
Therefore,
\begin{equation}
H_x=\frac{1}{\eta_0}\left(\beta E_z-\gamma E_y\right),
\end{equation}
\begin{equation}
H_y=\frac{1}{\eta_0}\left(\gamma E_x-\alpha E_z\right),
\end{equation}
and
\begin{equation}
H_z=\frac{1}{\eta_0}\left(\alpha E_y-\beta E_x\right).
\end{equation}
Since \(E_x=0\), the resulting magnetic components are
\begin{equation}
H_x=\frac{1}{\eta_0}\left(\beta E_z-\gamma E_y\right),
\qquad
H_y=-\frac{\alpha E_z}{\eta_0},
\qquad
H_z=\frac{\alpha E_y}{\eta_0}.
\end{equation}

Finally, all field components are transformed back to the source plane:
\begin{equation}
E_x(x,y),E_y(x,y),E_z(x,y)
=
\mathcal{F}^{-1}\{E_x,E_y,E_z\},
\end{equation}
\begin{equation}
H_x(x,y),H_y(x,y),H_z(x,y)
=
\mathcal{F}^{-1}\{H_x,H_y,H_z\}.
\end{equation}
The resulting FDTD source therefore injects the BPM scalar field as a forward-propagating vector electromagnetic field:
\begin{equation}
u_0(x,y)
\longrightarrow
\left[
E_x,E_y,E_z,H_x,H_y,H_z
\right].
\end{equation}
This construction is more physically consistent than assigning only \(E_y=u_0\) and \(H_x=-u_0/\eta_0\), because each propagating angular component receives the corresponding electromagnetic field components required by Maxwell's equations.

Although the source field is specified from a \(64\times64\) BPM input, the FDTD simulation is not performed on a \(64\times64\) grid. The custom source and all dielectric structures are interpolated onto the Yee grid used by the FDTD solver. In our simulations, the global mesh satisfies a minimum spatial resolution of approximately six cells per wavelength,
\begin{equation}
\Delta_{\mathrm{FDTD}}\lesssim \frac{\lambda}{6},
\end{equation}
and a finer mesh override is applied in the diffractive design region with a characteristic resolution of approximately
\begin{equation}
\Delta_{\mathrm{design}}\approx \frac{\lambda}{10}.
\end{equation}
For the visible-wavelength validation at \(\lambda=532~\mathrm{nm}\), the BPM pixel pitch is \(266~\mathrm{nm}\), corresponding to \(\lambda/2\). Therefore, each BPM pixel is represented by multiple FDTD cells, and the FDTD monitor outputs contain more spatial samples than the BPM detector arrays. When quantitative pixel-wise comparisons are required, the fields are placed on a common grid by either averaging the high-resolution FDTD intensity over BPM pixel regions or interpolating the BPM intensity onto the native FDTD monitor grid.

\subsection{Field evolution and sample results}
\label{app:fdtd_field_evolution}

\begin{figure*}[htbp]
    \centering
    \includegraphics[width=\textwidth]{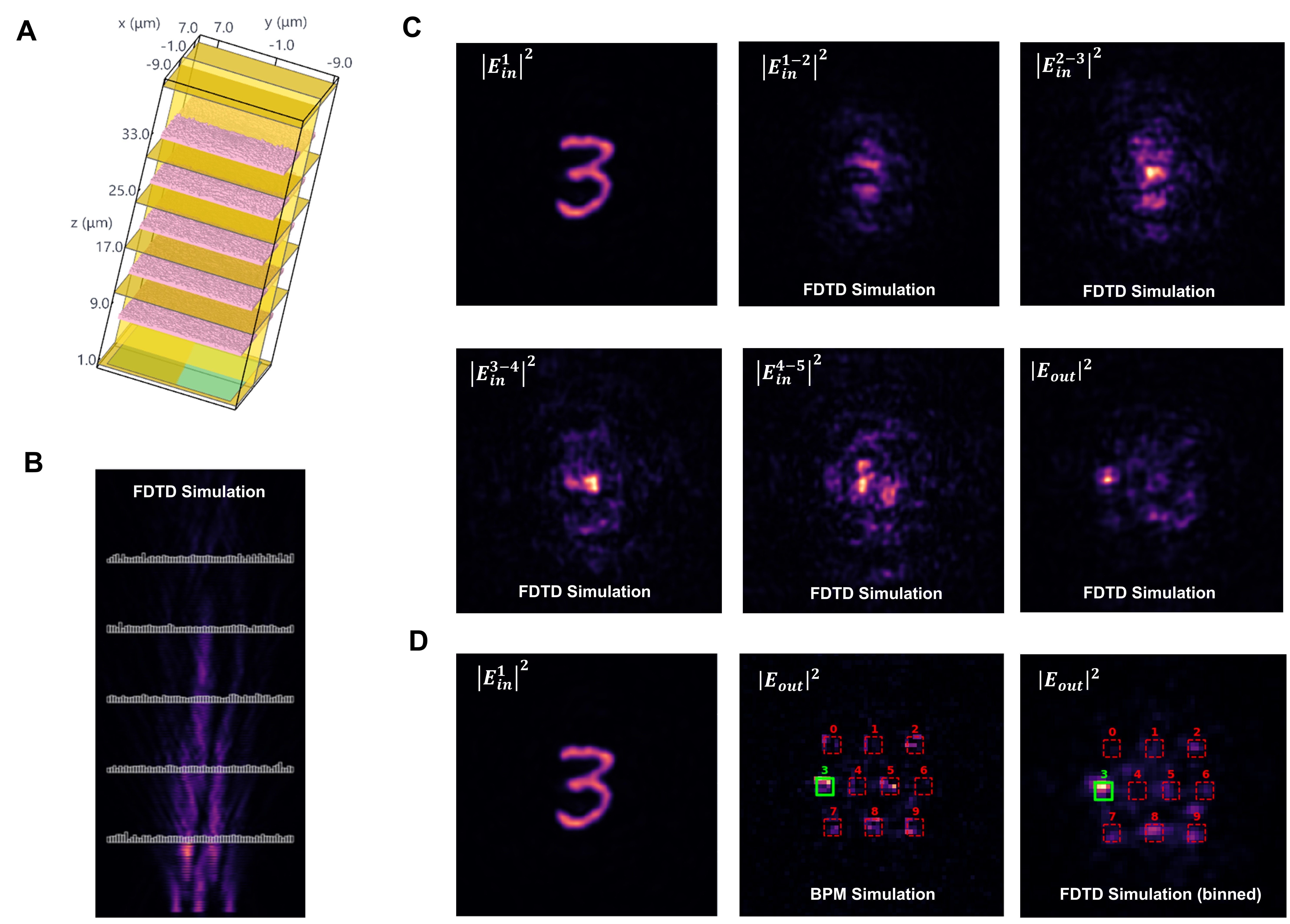}
    \caption{
    \textbf{Full-wave FDTD field evolution for a representative classification example.}
    \textbf{(A)} Reconstructed finite-thickness diffractive stack used in the FDTD simulation. The learned height maps are implemented as volumetric dielectric structures rather than zero-thickness phase masks.
    \textbf{(B)} Longitudinal FDTD field slice through the diffractive stack, showing how the optical field evolves along the propagation direction as it interacts with the finite-thickness layers.
    \textbf{(C)} Transverse intensity monitors at the input, between successive diffractive layers, and at the detector plane. The optical field progressively changes from the input digit pattern into a class-discriminative energy distribution. This illustrates that the classifier does not preserve the input image; instead, it redirects optical energy toward the target detector region.
    \textbf{(D)} Detector-plane comparison for the same sample. The BPM detector intensity and the binned FDTD detector intensity are shown with the class detector regions overlaid. The FDTD result is averaged onto the BPM pixel grid for comparison because the FDTD solver uses a finer electromagnetic mesh than the BPM model.
    }
    \label{fig:fdtd_classification_example}
\end{figure*}

Figure~\ref{fig:fdtd_classification_example} shows a representative full-wave FDTD validation example for the classification task. The purpose of this example is twofold: first, to visualize how the optical field evolves through the finite-thickness diffractive stack, and second, to compare the final detector-plane response with the corresponding BPM prediction. The learned height maps are reconstructed as finite-thickness dielectric layers in the FDTD solver, and the input field is injected using the Maxwell-consistent angular-spectrum source described above.

In the classification example, the field does not remain a simple image-like pattern after entering the diffractive stack. Instead, the diffractive layers progressively redistribute the optical energy across the transverse plane. The intermediate monitors show that the field is shaped through multiple stages: the input pattern is first diffracted by the early layers, then the energy is gradually concentrated toward the class-specific detector regions as it propagates through the remaining layers. This behavior is expected for a detector-based $D^2$NN classifier, because the network is not trained to preserve the input image at the output. Rather, it is trained to route optical power toward the detector patch corresponding to the target class. Therefore, the intermediate fields can appear increasingly non-image-like while still representing successful optical computation.

The detector-plane comparison in Fig.~\ref{fig:fdtd_classification_example}D shows the BPM output and the binned FDTD output for the same physical height map. The FDTD field is originally computed on a finer Yee grid, while BPM uses the coarser diffractive-pixel grid. For visual and quantitative comparison, the FDTD detector intensity is averaged over the corresponding BPM pixel regions. This comparison tests whether the BPM forward model predicts the same task-level detector response as the full-wave solver when both simulate the same finite-thickness structure.

\begin{figure*}[htbp]
    \centering
    \includegraphics[width=\textwidth]{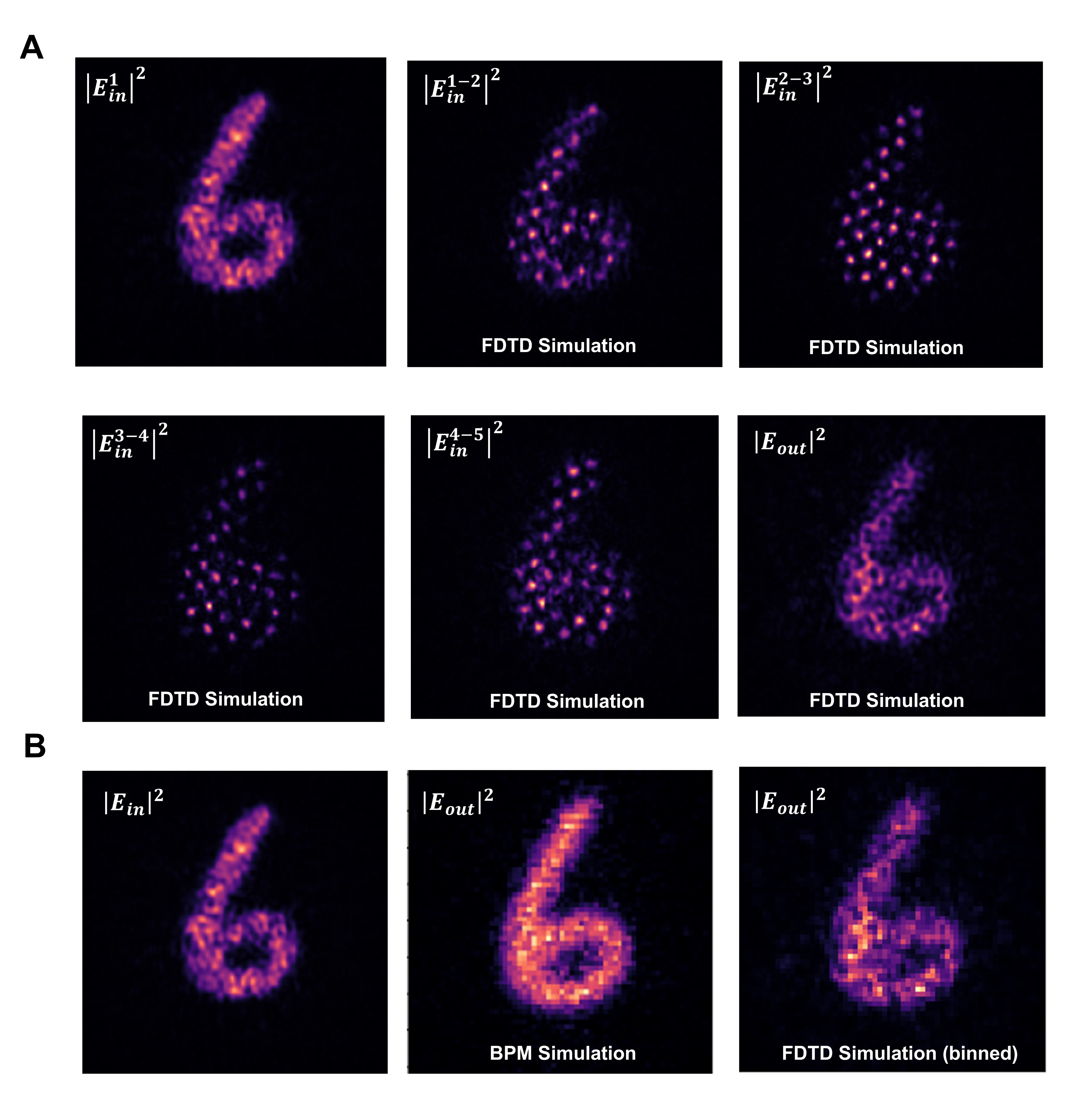}
    \caption{
    \textbf{Full-wave FDTD field evolution for a representative imaging example.}
    \textbf{(A)} Transverse intensity monitors through the finite-thickness diffractive stack. Unlike the classification task, the imaging task aims to preserve the input spatial structure at the detector plane, so the output remains image-like after propagation through the diffractive layers.
    \textbf{(B)} Output comparison for the same sample. The target/input image, BPM detector-plane reconstruction, and binned FDTD detector-plane reconstruction are shown. The FDTD output is averaged onto the BPM grid for direct comparison. The similarity between the BPM and FDTD reconstructions provides full-wave evidence that the BPM-trained volumetric structure preserves its intended imaging functionality.
    }
    \label{fig:fdtd_imaging_example}
\end{figure*}

Figure~\ref{fig:fdtd_imaging_example} shows an analogous FDTD validation example for the coherent imaging task. In contrast to classification, the imaging task is trained to preserve or reconstruct the spatial structure of the input at the output plane. Therefore, the detector-plane intensity should remain image-like. The intermediate FDTD monitors show that the field is still modified by propagation through the finite-thickness layers, but the final output retains the target image structure. This provides a complementary validation setting: classification tests whether the network routes energy to the correct detector region, while imaging tests whether the network preserves spatial information through the finite-thickness optical system.

The output comparison in Fig.~\ref{fig:fdtd_imaging_example}B shows the target/input image, the BPM detector-plane reconstruction, and the binned FDTD detector-plane reconstruction. Agreement between the BPM and FDTD outputs supports the use of the BPM model as a physically meaningful approximation for the finite-thickness structure. At the same time, the FDTD result verifies that the learned volumetric height maps preserve the intended imaging function under a Maxwell-equation-based solver.

\begin{figure*}[htbp]
    \centering
    \includegraphics[width=0.85\textwidth]{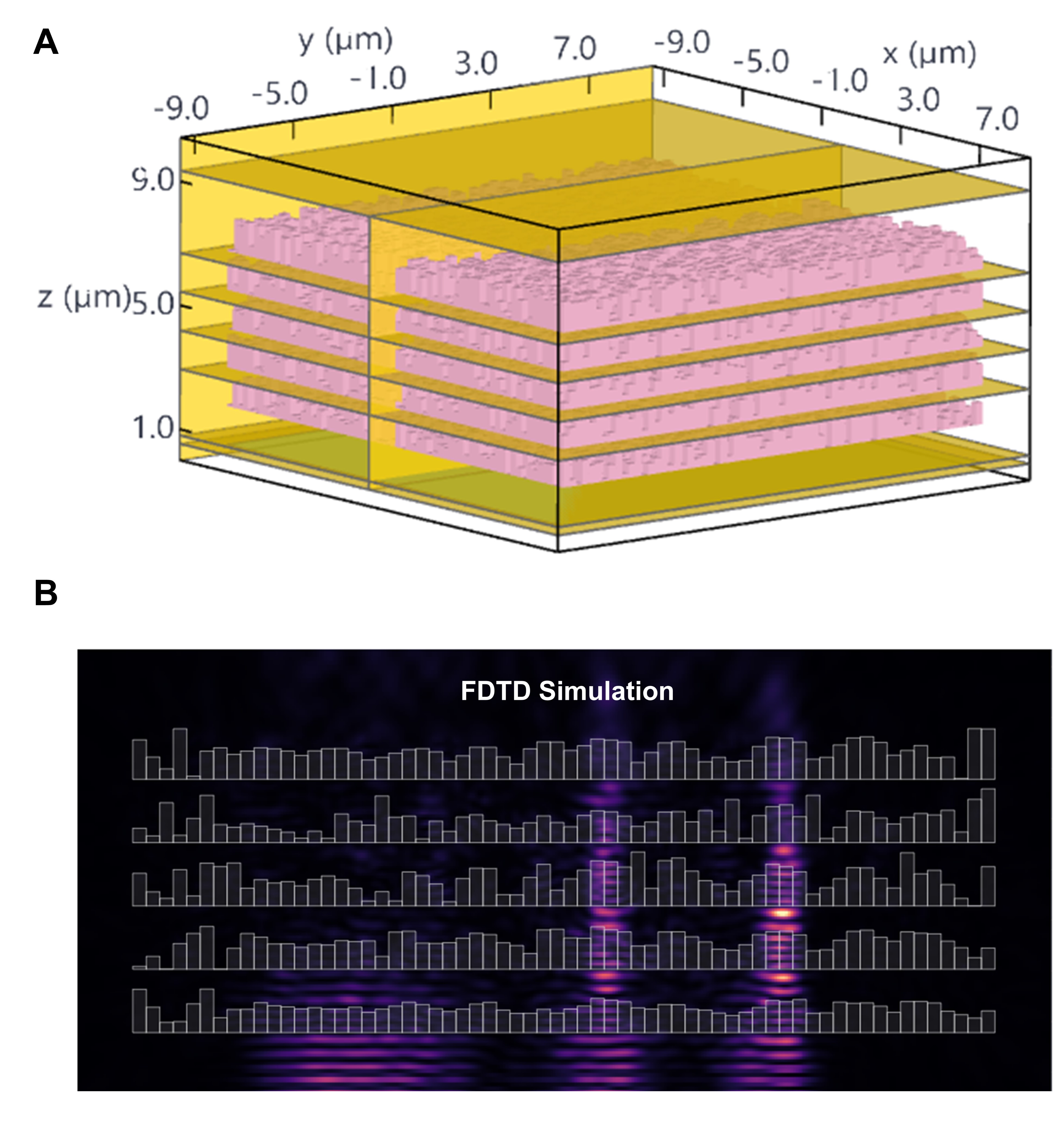}
    \caption{
    \textbf{Finite-thickness FDTD geometry and longitudinal field propagation.}
    \textbf{(A)} Three-dimensional view of the reconstructed diffractive stack used in the FDTD solver. The learned height maps are implemented as volumetric dielectric structures with nonzero axial thickness.
    \textbf{(B)} Longitudinal FDTD intensity slice through the structure. The field evolves continuously across the finite-thickness layers and free-space gaps, demonstrating that the physical device cannot be fully described as a cascade of infinitesimally thin phase screens. This supports the need for thickness-aware volumetric training.
    }
    \label{fig:fdtd_block_view}
\end{figure*}

Figure~\ref{fig:fdtd_block_view} provides a closer view of the finite-thickness FDTD geometry and the longitudinal field distribution. The geometry panel shows that each learned diffractive layer is represented as a physical height profile with finite axial extent. The longitudinal field slice shows that the optical field evolves continuously through and between the layers rather than undergoing instantaneous phase modulation at isolated planes. This visualization directly illustrates the physical effect targeted by the proposed BPM-D$^2$NN model: the diffractive layer is not merely a mathematical transmission mask, but a finite propagation volume.

\newpage
\section{Connection Between ASM Propagation, Evanescent Components, and the Neuron-Size Limit}

The angular spectrum method (ASM) provides a direct way to understand why the conventional diffraction-angle formula is no longer physically meaningful when the neuron size becomes smaller than $\lambda/2$. Consider an optical field $U(x,y,0)$ at the input plane. In ASM, this field is decomposed into its spatial-frequency components by a two-dimensional Fourier transform:
\begin{equation}
U(x,y,0)
\quad \longleftrightarrow \quad
\hat{U}(f_x,f_y,0),
\end{equation}
where $f_x$ and $f_y$ are the transverse spatial frequencies. In free space, each spatial-frequency component propagates over a distance $z$ by multiplication with the ASM transfer function:
\begin{equation}
H(f_x,f_y;z)
=
\exp\left[
i\frac{2\pi}{\lambda}z
\sqrt{1-\lambda^2 f_x^2-\lambda^2 f_y^2}
\right],
\end{equation}
where $\lambda$ is the free-space wavelength. Defining the free-space wavenumber as
\begin{equation}
k=\frac{2\pi}{\lambda},
\end{equation}
and writing the square-root term as
\begin{equation}
\gamma(f_x,f_y)
=
\sqrt{1-\lambda^2 f_x^2-\lambda^2 f_y^2},
\end{equation}
the propagation kernel can be written compactly as
\begin{equation}
H(f_x,f_y;z)=\exp\left(ikz\gamma\right).
\end{equation}

For propagating components, the square-root argument is non-negative:
\begin{equation}
1-\lambda^2 f_x^2-\lambda^2 f_y^2 \geq 0.
\end{equation}
In this case, $\gamma$ is real, and therefore
\begin{equation}
H(f_x,f_y;z)=\exp(ikz\gamma)
\end{equation}
is a pure phase factor with unit magnitude. The component propagates through free space and contributes to normal diffraction and inter-layer optical connectivity.

However, if
\begin{equation}
1-\lambda^2 f_x^2-\lambda^2 f_y^2 < 0,
\end{equation}
then the square-root term becomes imaginary. Using analytic continuation, we write
\begin{equation}
\gamma(f_x,f_y)
=
i\sqrt{\lambda^2 f_x^2+\lambda^2 f_y^2-1}.
\end{equation}
Let
\begin{equation}
\alpha
=
\sqrt{\lambda^2 f_x^2+\lambda^2 f_y^2-1}.
\end{equation}
Then
\begin{equation}
\gamma=i\alpha.
\end{equation}
Substituting this into the ASM transfer function gives
\begin{equation}
H(f_x,f_y;z)
=
\exp(ikz i\alpha),
\end{equation}
which simplifies to
\begin{equation}
H(f_x,f_y;z)
=
\exp(-kz\alpha).
\end{equation}
Thus, the component is no longer a propagating wave. Instead, it decays exponentially with propagation distance $z$. These are evanescent components, which remain confined near the diffractive structure and do not propagate efficiently to the next $D^2$NN layer.

This propagation limit can be connected directly to the finite neuron size $d_f$. In a pixelated diffractive layer, each neuron represents one spatial sample of the optical modulation function. Therefore, the neuron size $d_f$ also defines the sampling pitch of the diffractive layer. From the Nyquist sampling condition, a sampled spatial pattern can represent spatial frequencies only up to one half of the sampling frequency. Since the sampling frequency associated with the pixelated layer is $1/d_f$, the largest non-aliased transverse spatial frequency is approximately
\begin{equation}
f_{\max}\approx\frac{1}{2d_f}.
\end{equation}
Equivalently, the smallest spatial period that can be represented by the pixelated layer requires at least two neurons. Thus,
\begin{equation}
\Lambda_{\min}\approx 2d_f,
\end{equation}
and since spatial frequency is the inverse of spatial period,
\begin{equation}
f_{\max}\approx\frac{1}{\Lambda_{\min}}=\frac{1}{2d_f}.
\end{equation}

For free-space propagation, the angular spectrum method requires each propagating spatial-frequency component to lie inside the propagating spectral support:
\begin{equation}
\lambda^2 f_x^2+\lambda^2 f_y^2 \leq 1.
\end{equation}
This condition defines a circular propagating region in the two-dimensional spatial-frequency plane. To recover the conventional maximum half-cone diffraction-angle relation, we consider the one-dimensional limiting case with $f_y=0$ and $f_x=f_{\max}$. The propagation condition then becomes
\begin{equation}
\lambda f_{\max}\leq 1.
\end{equation}
Substituting $f_{\max}\approx 1/(2d_f)$ gives
\begin{equation}
\lambda\left(\frac{1}{2d_f}\right)\leq 1,
\end{equation}
or equivalently,
\begin{equation}
\frac{\lambda}{2d_f}\leq 1.
\end{equation}
Therefore, the neuron size must satisfy
\begin{equation}
d_f \geq \frac{\lambda}{2}.
\end{equation}

When
\begin{equation}
d_f < \frac{\lambda}{2},
\end{equation}
the maximum sampled spatial frequency associated with the pixelated neuron satisfies
\begin{equation}
f_{\max}>\frac{1}{\lambda}.
\end{equation}
These spatial-frequency components exceed the free-space propagating limit and make the ASM longitudinal propagation factor imaginary:
\begin{equation}
1-\lambda^2 f_x^2-\lambda^2 f_y^2 < 0.
\end{equation}
Consequently, $\gamma$ becomes imaginary and the corresponding field components become evanescent rather than propagating. These evanescent components decay exponentially with propagation distance and therefore do not contribute to far-field inter-layer propagation.

This also explains why the commonly used maximum half-cone diffraction-angle formula
\begin{equation}
u_{\max}=\sin^{-1}\left(\frac{\lambda}{2d_f}\right)
\end{equation}
is no longer physically meaningful when $d_f<\lambda/2$. This relation follows from the plane-wave angular spectrum condition
\begin{equation}
\sin u = \lambda f_x.
\end{equation}
Using the maximum sampled spatial frequency $f_x\approx 1/(2d_f)$, we obtain
\begin{equation}
\sin u_{\max}\approx\frac{\lambda}{2d_f}.
\end{equation}
However, if $d_f<\lambda/2$, then
\begin{equation}
\frac{\lambda}{2d_f}>1,
\end{equation}
which would require
\begin{equation}
\sin u_{\max}>1.
\end{equation}
No real propagating angle can satisfy this condition. In the ASM description, this same regime corresponds to an imaginary longitudinal propagation factor and exponential evanescent decay. Therefore, below the $\lambda/2$ neuron-size limit, reducing $d_f$ does not continue to increase far-field diffraction connectivity. Instead, the additional high-spatial-frequency content becomes evanescent and does not contribute to free-space inter-layer propagation in a conventional $D^2$NN.

\section{Derivation of the Inter-Layer Connectivity Condition}

The above neuron-size argument explains when a spatial-frequency component can propagate in free space. We now connect this propagation limit to the geometric connectivity condition used for visible-wavelength $D^2$NN design. In a free-space $D^2$NN, the field diffracted by each neuron should spread sufficiently over the next diffractive layer. Otherwise, each neuron only communicates with a small local region of the next layer, reducing the effective optical connectivity of the network.

Let $d_f$ denote the lateral size of one diffractive neuron, and let $\lambda$ denote the operating wavelength. Following the maximum half-cone diffraction-angle approximation, the largest propagating diffraction angle produced by one neuron can be written as
\begin{equation}
u_{\max}=\sin^{-1}\left(\frac{\lambda}{2d_f}\right).
\end{equation}
This expression follows from the spatial-frequency relation
\begin{equation}
\sin u=\lambda f_x,
\end{equation}
together with the maximum representable spatial frequency of a pixelated neuron,
\begin{equation}
f_{\max}=\frac{1}{2d_f}.
\end{equation}
Substituting $f_x=f_{\max}$ gives
\begin{equation}
\sin u_{\max}=\lambda f_{\max}=\frac{\lambda}{2d_f}.
\end{equation}
Therefore,
\begin{equation}
u_{\max}=\sin^{-1}\left(\frac{\lambda}{2d_f}\right).
\end{equation}

Now consider two adjacent diffractive layers separated by an axial propagation distance $D$. If light from one neuron propagates with maximum half-angle $u_{\max}$, then its lateral spreading radius on the next layer is obtained from simple geometric optics. The axial distance is $D$, the half-angle is $u_{\max}$, and the lateral radius of the diffraction cone is $R$. Therefore,
\begin{equation}
\tan(u_{\max})=\frac{R}{D}.
\end{equation}
Rearranging gives
\begin{equation}
R=D\tan(u_{\max}).
\end{equation}

Next, assume that each diffractive layer is square and contains $N$ neurons. Each neuron has lateral area $d_f^2$. Therefore, the total area of one layer is
\begin{equation}
A=N d_f^2.
\end{equation}
If the layer is square with side length $w$, then
\begin{equation}
w^2=A=N d_f^2.
\end{equation}
Taking the square root gives
\begin{equation}
w=\sqrt{N d_f^2}=\sqrt{N}d_f.
\end{equation}

For full inter-layer connectivity, the diffraction cone from a neuron should be wide enough to cover the next diffractive layer. Therefore, the lateral diffraction radius must satisfy
\begin{equation}
R\geq w.
\end{equation}
Substituting $R=D\tan(u_{\max})$ and $w=\sqrt{N}d_f$ gives
\begin{equation}
D\tan(u_{\max})\geq \sqrt{N}d_f.
\end{equation}
Using the expression for the maximum diffraction angle,
\begin{equation}
u_{\max}=\sin^{-1}\left(\frac{\lambda}{2d_f}\right),
\end{equation}
we obtain
\begin{equation}
D\tan\left[\sin^{-1}\left(\frac{\lambda}{2d_f}\right)\right]\geq \sqrt{N}d_f.
\end{equation}

This expression can be simplified using the trigonometric identity
\begin{equation}
\tan(\sin^{-1}x)=\frac{x}{\sqrt{1-x^2}}.
\end{equation}
Let
\begin{equation}
x=\frac{\lambda}{2d_f}.
\end{equation}
Then
\begin{equation}
\tan\left[\sin^{-1}\left(\frac{\lambda}{2d_f}\right)\right]=\frac{\lambda/(2d_f)}{\sqrt{1-\lambda^2/(4d_f^2)}}.
\end{equation}
Simplifying the denominator gives
\begin{equation}
\sqrt{1-\frac{\lambda^2}{4d_f^2}}=\sqrt{\frac{4d_f^2-\lambda^2}{4d_f^2}}=\frac{\sqrt{4d_f^2-\lambda^2}}{2d_f}.
\end{equation}
Therefore,
\begin{equation}
\tan\left[\sin^{-1}\left(\frac{\lambda}{2d_f}\right)\right]=\frac{\lambda}{\sqrt{4d_f^2-\lambda^2}}.
\end{equation}

Substituting this into the full-connectivity condition gives
\begin{equation}
D\frac{\lambda}{\sqrt{4d_f^2-\lambda^2}}\geq \sqrt{N}d_f.
\end{equation}
Solving for $D$ gives
\begin{equation}
D\geq \frac{\sqrt{N}d_f\sqrt{4d_f^2-\lambda^2}}{\lambda}.
\end{equation}
Equivalently, this can be written as
\begin{equation}
D\geq \sqrt{N}d_f\sqrt{\frac{4d_f^2}{\lambda^2}-1}.
\end{equation}

This is the inter-layer connectivity condition. It shows that the connectivity of a free-space $D^2$NN depends on the wavelength $\lambda$, neuron size $d_f$, number of neurons $N$, and inter-layer spacing $D$. For a fixed neuron size and layer size, reducing the wavelength decreases the maximum diffraction angle and therefore reduces the diffraction radius $R$. To restore full connectivity at shorter wavelengths, one must either reduce the neuron size $d_f$, increase the inter-layer spacing $D$, or reduce the number of neurons $N$.

This relation explains the central visible-wavelength design tradeoff. When $\lambda$ is reduced from the terahertz regime to the visible regime, the maximum diffraction angle becomes much smaller for the same neuron size. Maintaining a large diffraction angle therefore requires reducing $d_f$ toward the nanoscale, which increases fabrication difficulty. Alternatively, one can keep $d_f$ in a more fabricable micrometer-scale range and increase $D$, allowing the diffracted light more distance to spread before reaching the next layer. Thus, visible-wavelength $D^2$NNs face a wavelength--neuron-size--spacing tradeoff: compact devices require very small neurons, whereas easier-to-fabricate larger neurons require longer propagation distances.

% \bibliography{sample}

\end{document}